\documentclass[]{spie}  
\usepackage[]{graphicx}
\usepackage[square,comma,sort,numbers]{natbib}
\usepackage{comment}
\usepackage{xcolor}
\usepackage{amssymb}
\usepackage{array}
\usepackage{subcaption}
\usepackage{rotating}
\usepackage{wrapfigure}
\usepackage[section]{placeins}

\title{Towards controlled Fizeau observations with the Large Binocular Telescope\footnote{The LBT is an international collaboration among institutions in the United States, Italy and Germany. LBT Corporation partners are: The University of Arizona on behalf of the Arizona university system; Istituto Nazionale di Astrofisica, Italy; LBT Beteiligungsgesellschaft, Germany, representing the Max-Planck Society, the Astrophysical Institute Potsdam, and Heidelberg University; The Ohio State University, and The Research Corporation, on behalf of The University of Notre Dame, University of Minnesota and University of Virginia.}} 

\author{Eckhart Spalding\supit{a}, Phil Hinz\supit{a}, Steve Ertel\supit{a}, Erin Maier\supit{a}, Jordan Stone\supit{a}
\skiplinehalf
\supit{a}Steward Observatory, University of Arizona, 933 Cherry Ave., Tucson, AZ, 85721, USA
}

\authorinfo{Further author information: (Send correspondence to E.S.)\\E.S.: E-mail: spalding@email.arizona.edu}

\begin{document} 
  \maketitle 

\begin{abstract}
The Large Binocular Telescope Interferometer (LBTI) can perform Fizeau interferometry in the focal plane, which accesses spatial information out to the LBT's full 22.7-m edge-to-edge baseline. This mode has previously been used to obtain science data, but has been limited to observations where the optical path difference (OPD) between the two beams is not controlled, resulting in unstable fringes on the science detectors. To maximize the science return, we are endeavoring to stabilize the OPD and tip-tilt variations and make the LBTI Fizeau mode optimized and routine. Here we outline the optical configuration of LBTI's Fizeau mode and our strategy for commissioning this observing mode.
\end{abstract}



\section{INTRODUCTION}
\label{sec:intro}  

The facility which was later to become the LBT had interferometry directly in its strategic sights since the facility's earliest conceptual designs in the 1980s \cite{woolf1983versatile}. It was recognized that a large, $>10$ m class telescope optimized for the infrared would unlock a host of science cases, but that manufacturing and handling limitations precluded the construction of a giant, continuous mirror. 

Consequently, and based on experience with the co-phased Multiple Mirror Telescope (MMT), the LBT aperture was ultimately split into two 8.4 m mirrors separated by 6 m \cite{mccarthy1988interferometry,hill1994strategy}. The inner separation of 6 m between the two primary mirrors was ``a compromise between maximum angular resolution (i.e., maximum baseline) and continuity of spatial frequency response" \cite{mccarthy1988interferometry}. In other words, the shortest baselines which stretch from one mirror to the other are degenerate with the longest baselines achievable within a single mirror, which allows continuity in  the LBT aperture's sampling of frequency space of observed targets (Table \ref{table:fiz_img_analy}). 

Today, the LBT's discontinuous aperture has the longest baseline of any other astronomical telescope on a single mount. The LBT is the only infrared-optimized telescope in the world with an achievable resolution on the scale of future extremely large telescopes (ELTs), and it will be in a unique position to corner ELT science cases until the first ELTs are operational in the mid-2020s. Indeed, those first ELTs (the GMT and E-ELT) will be southern facilities, and the northern TMT is projected to come online in the late 2020s. 

Since atmospheric aberrations destroy angular information content over large baselines, infrared imaging interferometry requires a high-end adaptive optics system. As of spring 2018, the LBT/LBTI adaptive optics (AO) system samples the wavefront deformation over a grid of up to 30$\times$30 over the aperture, and sends compensating signals at up to $\sim$1 kHz to a deformable secondary mirror. The use of the secondary mirror as the deformable mirror acts to improve throughput and remove infrared background emission \cite{lloyd2000thermal}. 

The LBT Interferometer (LBTI) instrument sits between the primary mirrors and can access the LBT's dual-aperture baselines through interferometric synthesis. After the light from the primary mirrors hits the deformable secondary mirrors, tertiary mirrors guide the f/15 beams into the LBTI's cryogenically-cooled Universal Beam Combiner (UBC) \cite{hinz2004large}. The UBC guides the beams within a f/15-envelope to a common focus, after which the light is sent to either of two science channels: LMIRcam \cite{skrutskie2010large,leisenring2012sky} (1-5 $\mu$m) and/or NOMIC \cite{hoffmann2014operation} (currently 8-14 $\mu$m). A third channel, Phasecam \cite{defrere2014co}, is a fringe-tracking detector sensitive to $H$- and $K_{S}$-bands. 

\begin{figure}
\begin{center}
\includegraphics[height=9cm]{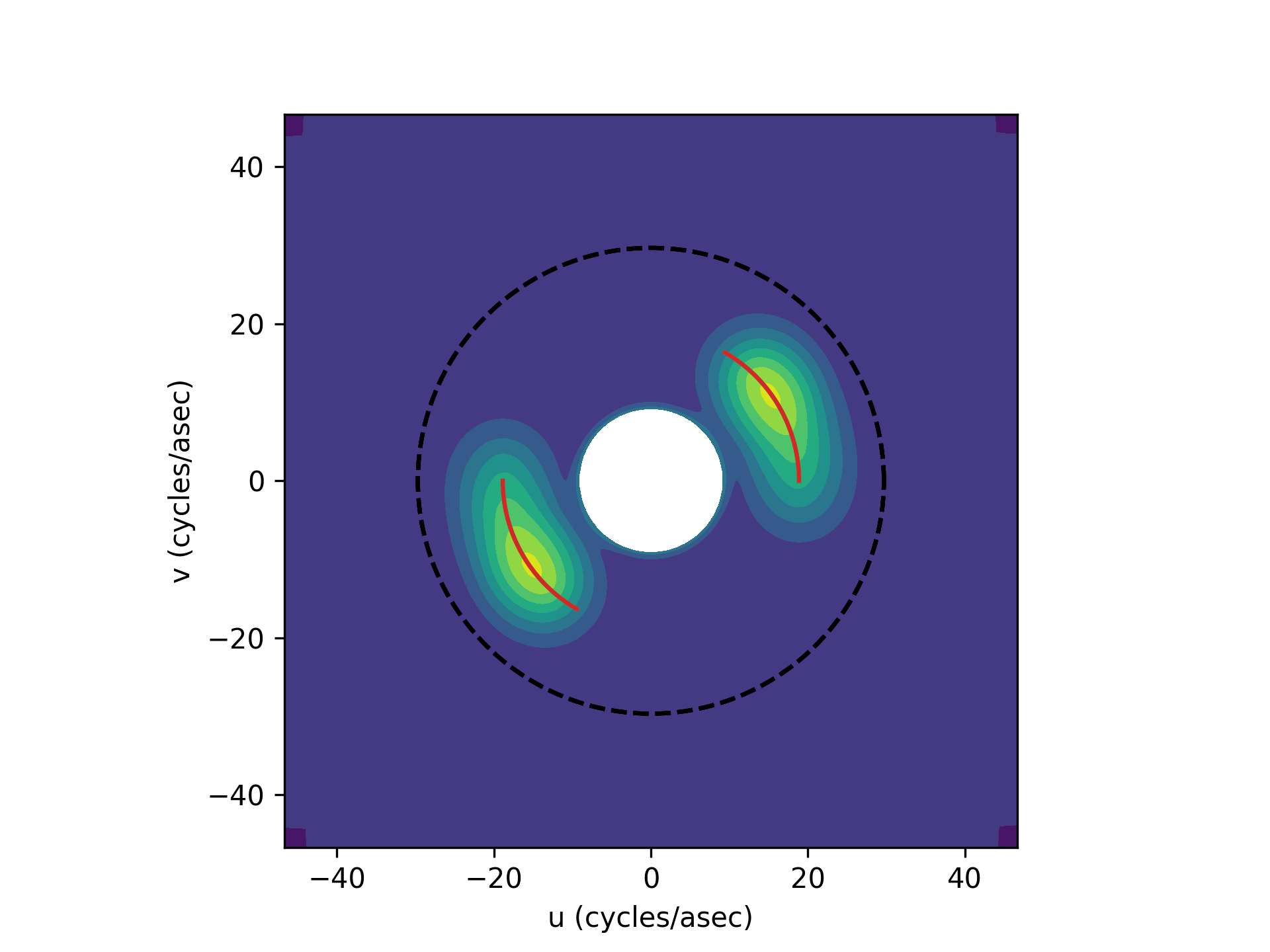}
\caption[]{A heuristic representation of the $uv$ coverage of an observation at 3.7 $\mu$m with 60 degrees of rotation. The curved, red solid lines indicate the $uv$ coverage if the observation were obtained in a nulling (co-axial) configuration. This representation is inadequate in Fizeau (multi-axial) mode, where the $uv$ coverage must be represented with a contour plot, here in linear scale. Note that the contours are not circularly symmetric because the speed of the on-sky parallactic angle rotation is not constant. (As in \cite{mccarthy1988interferometry}, we mask the low-frequency node to bring out detail in the higher frequencies.) The dashed line indicates the outer limit of obtainable frequency information, as imposed by the edge-to-edge baseline of the two primary mirrors.}
\label{fig:rot60deg}
\end{center}
\end{figure}

LBTI can observe in a number of high-contrast modes (see \cite{hinz18b} in these proceedings). These include two main interferometric modes: 1.), nulling with NOMIC, in which the beams are combined in a plane close to a pupil plane (i.e., co-axial beam combination) and a $\pi$ phase shift in one arm is used to place a destructive fringe over the central light source \cite{bracewell1978detecting}; or 2.) imaging (also known as ``Fizeau'') interferometry with LMIRcam and/or NOMIC, in which the beams are combined in a focal plane (i.e., multi-axial beam combination) and the sampled $uv$ is more extended in frequency space, making available more spatial information at continuous frequencies for image reconstruction. (See Fig. \ref{fig:rot60deg}.)

Nulling is advantageous for suppressing the bright light from host stars and for detecting faint, extended circumstellar material inside small fields-of-view. Since the pupils are overlapped, there is a one-to-one mapping of pairs of points between the two apertures at separations of 14.4 m. Thus nulling allows the extraction of angular resolutions equivalent to the 14.4 m center-to-center mirror separation. In Fizeau interferometry, the beam combination in the focal plane allows extraction of detail out to the 22.7 m edge-to-edge separation of the two mirrors.\footnote{This includes the stopping-down of the LBTI pupils by the undersized secondary mirrors.} This represents by far the largest optical or infrared Fizeau baseline in the world.

Until now, the great majority of LBTI interferometric observations have been in nulling mode as part of the Hunt for Observable Signatures of Terrestrial Systems (HOSTS) survey of exozodiacal dust around nearby ($<30$ pc) stars \cite{weinberger2015target,ertel2018hosts,ert18b}. This survey has been funded since 2001 by  NASA's Exoplanet Exploration Program (ExEP)\footnote{Part of the NASA Science Mission Directorate Astrophysics Division} in order to constrain the noise contribution of exozodiacal dust so as to inform designs of future space-based exoplanet direct imaging missions, particularly with regard to primary mirror sizes \cite{brown2005single,roberge2012exozodiacal}. HOSTS science observations began in the spring 2014 observing season and came to a successful end in spring 2018, setting the stage for a segue to commissioning of the Fizeau interferometric mode.

Our main goal in this article is to outline how we will implement a ``controlled'' Fizeau mode, whereby the single-aperture PSFs are kept overlapped and the fringes are actively kept stable, and at the center of the coherence envelope. Stable fringes will avoid phase smearing and will increase observing efficiency by making a larger proportion of detector readouts useable and by allowing longer integrations.

The technical potential for commissioning routine imaging interferometry in the infrared is scientifically timely. High-contrast, AO-corrected imaging interferometry with the LBT can unlock a host of cutting-edge science involving thermally-emissive structure at tiny angular scales, including the detailed imaging of protoplanetary and transition disks, binarity searches in crowded fields, and the probing of the distribution of material in the inner regions of exoplanetary systems. Among other targets, the Taurus-Auriga region of natal circumstellar disks, at 140 pc away, is clearly one of the most interesting. Other emerging facilities like ALMA will provide complementary data on natal exoplanetary systems (e.g., \cite{elmegreen2015optimizing}). With nulling in a highly commissioned state, controlled Fizeau is the next step in expanding LBTI's interferometric capabilities.

\begin{figure}
\begin{center}
\includegraphics[width=1.0\linewidth, trim={30cm, 18cm, 29cm, 26cm}, clip=True]{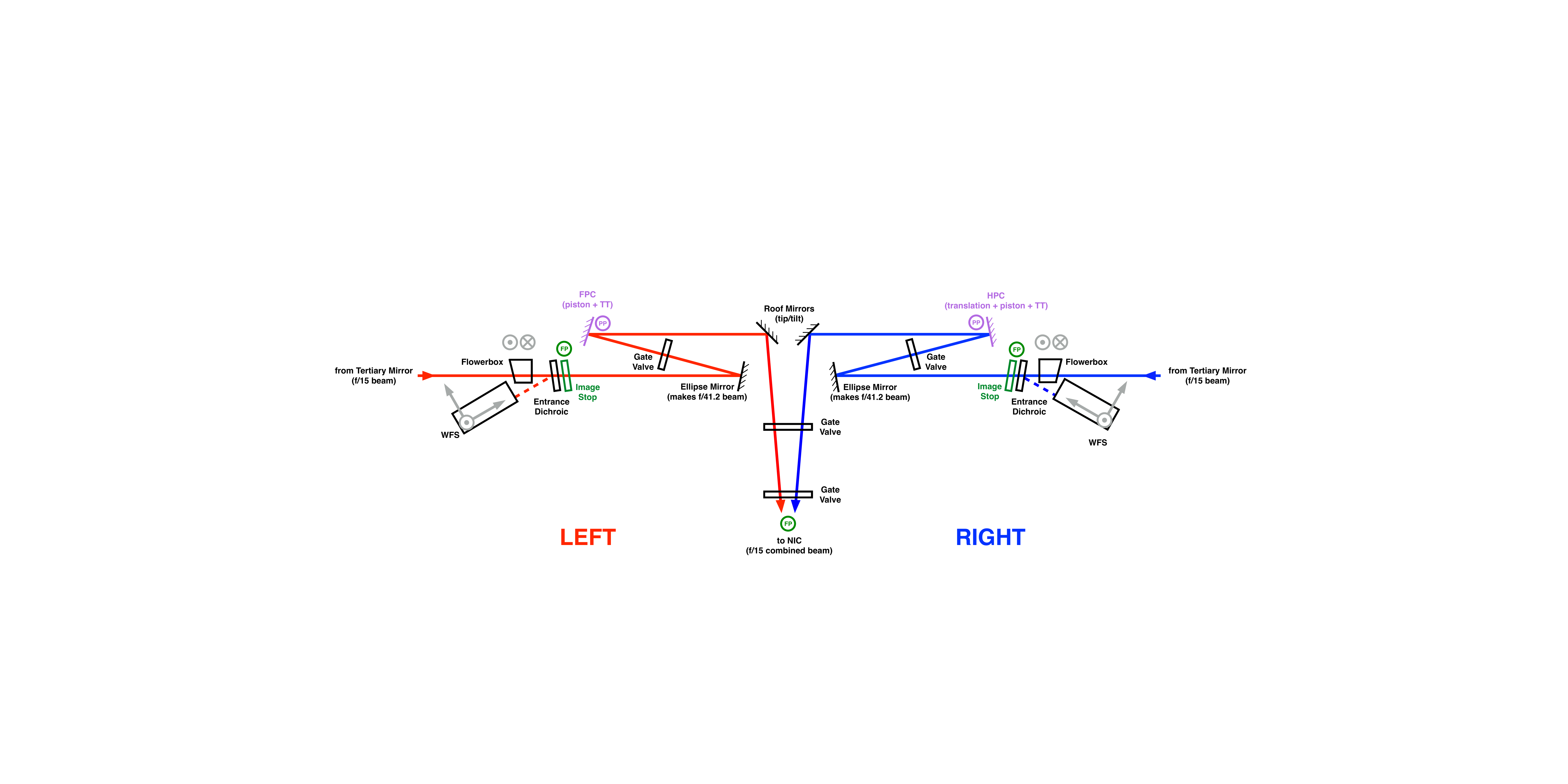}
\caption[]{A pencil-beam schematic of the Universal Beam Combiner (UBC), upstream of the NIC cryostat, , as of spring 2018. This shows elements and sources of pathlength and tip-tilt the beams encounter, as in Fig. \ref{fig:nic_diag}. Some liberty has been taken with the fold angles for clarity. (For more physical angles, see, for example, \cite{hinz2004large}.) Grey arrows (or arrowhead points and tails) show possible directions of movement of the wavefront sensors, and ``flowerboxes'' which hold a calibration laser source.}
\label{fig:ubc_diag}
\end{center}
\end{figure}

\begin{sidewaysfigure}
\begin{center}
\includegraphics[width=0.77\linewidth, trim={16cm, 0cm, 41cm, 0cm}, angle=0, clip=True]{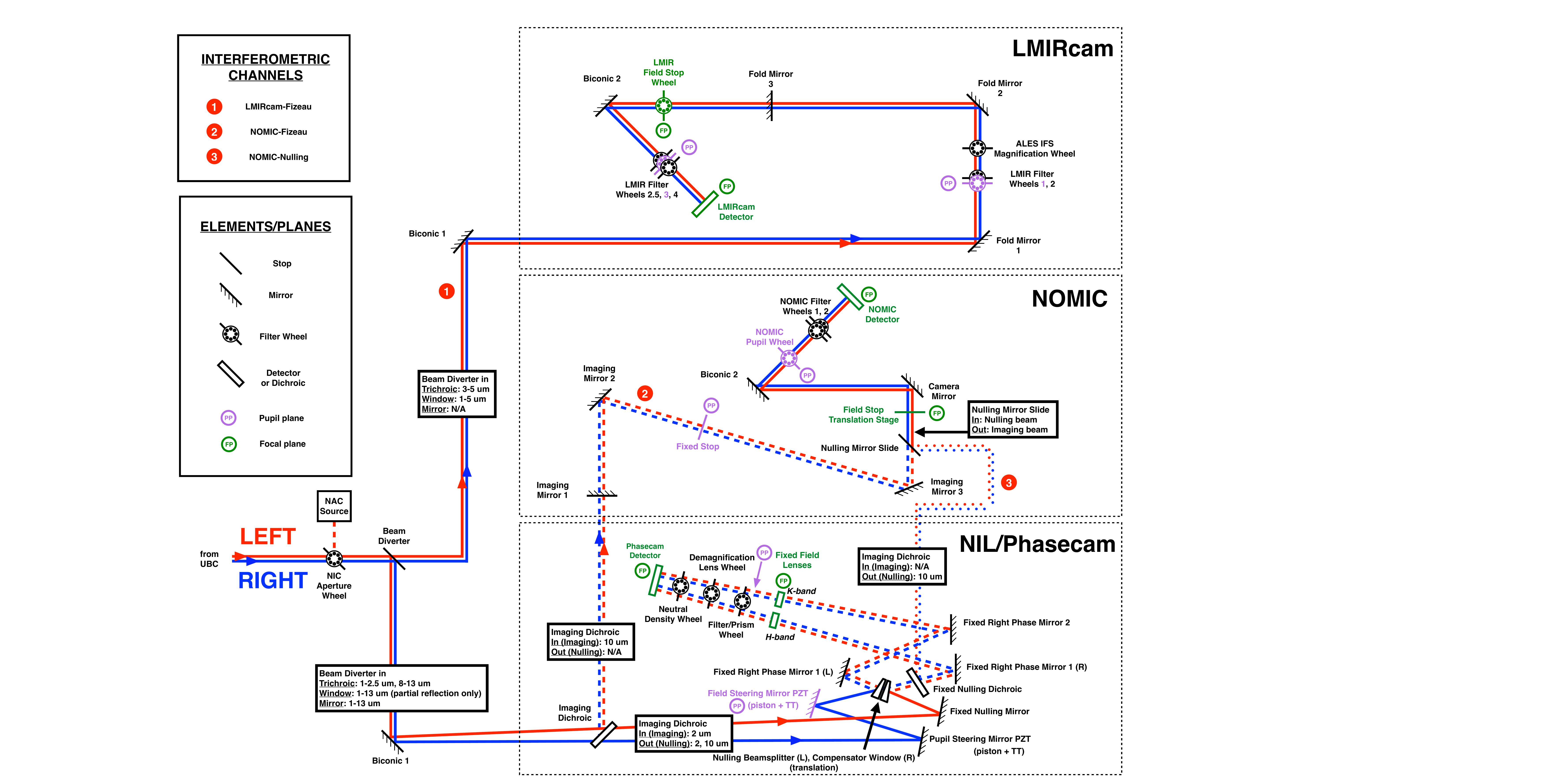} 
\vspace{0.3cm}
\caption[]{A schematic of the Nulling Infrared Camera (NIC) cryostat, which is downstream of the Universal Beam Combiner (UBC), as of spring 2018. This schematic is drawn to indicate all optical elements the beams encounter. All instrumental sources of alterable tip, tilt, and OPD are indicated, in parentheses under the relevant element: ``TT'' is tip/tilt; ``piston'' is micron-scale piezo piston (and hence OPD) variations; ``translation'' is millimeter-scale OPD by moving the element on a translation stage. (See Table \ref{table:sources_opd_tt}.) Elements in pupil planes are indicated with purple color and the `PP' symbol, and elements in focal planes with green color and `FP'. Each dotted box represents a physical level of the cryostat. Note that in closed Phasecam loop, the Phasecam detector image is actually closest to a pupil plane because of the Demag Lens Wheel. The science detectors can also image pupil planes with an appropriate lens in LMIRcam's filter wheel 4, or NOMIC's filter wheel 1.)}
\label{fig:nic_diag}
\end{center}
\end{sidewaysfigure}

\section{The state of Fizeau science with LBTI} 

\subsection{Prior work}
\label{subsec:nascent}

The LBT obtained its first seeing-limited Fizeau fringes in October 2010 in $N$-band, using the former MMT telescope instrument MIRAC4-BLINC while it was hooked up to the LBTI external structure for preparatory tests \cite{hinz2008nic,hinz2012first}. The LBT right and left adaptive secondary mirrors were installed in 2010 and 2011, and have been doing regular science since 2011 and 2012, respectively \cite{riccardi2010adaptive,hill2012large,brynnel2012commissioning}. LBTI (with the current NIC cryostat \cite{hinz2008nic}) obtained the first AO-corrected (but not phase-controlled) Fizeau images with LMIRcam and NOMIC in April and May 2012 \cite{hill2012large}. Finally, $K_{S}$-band phase-tracking with Phasecam was coupled to the nulling mode beginning in December 2013 \cite{wagner2014overview}. 


An LBT Fizeau PSF would ideally look very much like the simulated image in the top row of Table \ref{table:fiz_img_analy}, or more physically like that in Fig. \ref{fig:phys_psf}.\footnote{We should note that strictly speaking, the PSF is the target-independent instrument impulse response, which must be convolved with a perfect, instrument-independent image of the object $O$ to yield the image $I$ on the detector, or $I = PSF * O$. For simplicity, however, we use ``PSF'' loosely in this article to refer to the irradiance on the detector, because we only consider point source targets (i.e., delta functions). Naturally, our ultimate hope is to conduct Fizeau observations of objects considerably more interesting than delta functions.} The familiar bright and dark Airy rings of a diffraction-limited PSF from a single, circular aperture of 8.25 m (undersized from 8.4 m by the secondary mirror) is multiplied by a linear sinusoid from the offset between the two apertures. For a finite bandwidth, high-contrast fringes are only seen near an optical path difference (OPD) of zero between the two beams. The LBT's AO system routinely corrects the wavefronts in the beams to Strehls of $\gtrsim85\%$ at 4 $\mu$m \cite{bailey2014large}, but the two adaptive secondaries apply independent corrections which are blind to differential piston. There can still be strongly time-varying, atmospherically-induced differential piston (on the order of 10 $\mu$m on a timescale of a few minutes \cite{hill2013large}) or other aberrations between the two corrected wavefronts. 

Fringe phase in the PSF depends on the OPD between the two beams, and differential piston manifests itself in the LBT PSF as a rolling of the fringes through the Airy rings (Table \ref{table:fiz_img_analy}, row 4).  If the OPD veers further away from the center of the fringe packet, the visibility of the fringes decreases. At large enough OPD, the beams are essentially incoherent, and the illumination on the detector becomes that of a single Airy function with twice the amplitude of a single aperture (Table \ref{table:fiz_img_analy}, row 5). To avoid fringe smearing and remain in the center of the coherence envelope, the path difference between the beams must be actively controlled. 

Some early scientific work has been done in ``lucky'' Fizeau mode, involving detailed $M$-band imaging of the volcanically active Jovian moon Io \cite{leisenring2014fizeau,conrad2015spatially,conrad2016role,de2017multi}. A small handful of other targets have been Fizeau-imaged for testing purposes, including Vega \cite{hoffmann2014operation}, HD 184786 \cite{hill2013large}, CH Cyg, and a collection of stars in the Trapezium asterism, separated by several arcseconds to demonstrate wide-field co-phasing \cite{hinz2014commissioning}. However, without active phase control, time-varying differential OPD and tip-tilt instability remain, and cause fringes to jitter back-and-forth in the focal plane. Furthermore, short integration times are necessary to avoid phase smearing \cite{hinz2012first,sallum2017improved}. In the reduction phase, the ``lucky'' frames with centered fringes are sieved from the rest. In terms of the useful detector integration time compared to total integration time, the efficiency until now has been extremely low. Stated efficiencies in the literature range from 2-5 \% \cite{conrad2015spatially}, $\sim$5-10\% \cite{leisenring2014fizeau}, and $\sim$10-20\% \cite{conrad2016role}. These percentages do not include time that is lost 1.) during the readout of the necessarily short exposures to avoid fringe smearing (e.g., \cite{leisenring2014fizeau,conrad2015spatially}), and 2.) due to the manual nature of the observation, such as when members of the observing team have to manually tune a correction mirror at each telescope dither position, to find the center of the coherence envelope as best as possible.\footnote{Note that a bright resolved source like Io cannot be phase-controlled with Phasecam, because the extended nature of the source means that the interferometric visibility will be too low. Nevertheless, these low observing efficiencies illustrate what one would get by observing a point source without active phase control.}

Since a bright fringe passes through the PSF center for every wavelength of OPD, the fringes in the ``lucky'' frames are of unknown distance in OPD from the center of the coherence envelope, where the visibility is best. The fringes in ``lucky'' frames are from \textit{somewhere} near the center of the coherence envelope, but it is unknown \textit{how} near. A simple measurement of fringe positions from one image to the next only yields OPD modulo $\lambda$, rather than a true OPD (as measured from the center of the coherence envelope, or in the differential OPD from one frame to the next). There is a lot of room to move---depending on the transmission filter profile, there could be ten or so bright fringes in the coherence envelope (e.g., Fig. \ref{table:fiz_open_closed}). If the frames with centered fringes are co-added for higher signal-to-noise, the obtainable contrast is washed out through the mixture of different phases.

\subsection{The current technical problem}

Design of any strategy for controlling and optimizing the Fizeau imaging mode requires an understanding of the two beam paths. After reflecting off the primary mirrors, the atmospherically-aberrated wavefronts are, to a high approximation, flattened out by the deformable secondary mirrors. The light then reflects off the static telescope tertiary mirrors and towards the telescope bent Gregorian foci at the LBTI Universal Beam Combiner (UBC). A dichroic sends the visible light into the wavefront sensors while allowing the infrared light to proceed into the cryogenic UBC (Fig. \ref{fig:ubc_diag}) and on into the Nulling Infrared Camera (NIC) cryostat which contains the detectors (Fig. \ref{fig:nic_diag}). 

In nulling mode, the Phasecam (phase-tracking) and NOMIC (science) channels are common until they reach a nulling dichroic which sends the $H$- and $K_{S}$-band light to Phasecam and the $N$-band light to NOMIC. The detectors are both downstream of all (intentionally) adjustable sources of OPD, tip, and tilt. (See Table \ref{table:sources_opd_tt}.) 

\begin{table}
\begin{center}
\caption{Controllable sources of OPD, Tip, Tilt} 
\label{table:sources_opd_tt}
\begin{tabular}{| c | c | c | c | l |}
\hline
\textit{Element}	& 	\textit{Plane}	& \begin{tabular}{@{}c@{}}Upstream \\ of all \\detectors?\end{tabular}  & \begin{tabular}{@{}c@{}} \textit{Degree of freedom / range} \\ \textit{on sky at cryogenic temperatures}\end{tabular}	& \textit{Remarks / advantage} \\
 \hline	
\begin{tabular}{@{}c@{}}NIL beam  \\ combiner\end{tabular}	& 		$\approx$Pupil	& N	& OPD: 24 mm				&  \begin{tabular}{@{}l@{}}Co-alignment of coherence \\  envelopes during setup\end{tabular}  \\	
\hline
\begin{tabular}{@{}c@{}}Fast pathlength  \\ corrector (FPC)\end{tabular}			& 	Pupil 	& Y	&  \begin{tabular}{@{}c@{}}OPD: $2\times50$ $\mu$m (piezo) \\= 100 $\mu$m \\ Tip/tilt: $\approx\pm$5 asecs \end{tabular}	& \begin{tabular}{@{}l@{}}Automated removal of \\differential piston, 1 kHz \\(Phasecam loop) \end{tabular} \\	
\hline
\begin{tabular}{@{}c@{}}Hybrid pathlength  \\ corrector (HPC)\end{tabular}			& 	Pupil 	& Y	& \begin{tabular}{@{}c@{}}OPD: $2\times16$ mm (transl. stage)\\+$2\times50$ $\mu$m (piezo) $\approx$ 32.1 mm \\ Tip/tilt: $\approx\pm$5 asecs \end{tabular}	& \begin{tabular}{@{}l@{}}Manual scanning for \\coherence envelope \\during setup\end{tabular} \\	
\hline
\begin{tabular}{@{}c@{}}Pupil steering  \\ mirror (PSM)\end{tabular}			& 	-- 	& N	& \begin{tabular}{@{}c@{}}OPD: $2\times20$  $\mu$m =  40  $\mu$m \\ Tip/tilt: $\approx\pm$0.4 asecs \end{tabular}	& Semi-static\\
\hline
\begin{tabular}{@{}c@{}}Field steering  \\ mirror (FSM)\end{tabular}			& 	$\approx$Pupil 	& N	& \begin{tabular}{@{}c@{}}OPD: $2\times20$  $\mu$m =  40  $\mu$m \\ Tip/tilt: $\approx\pm$0.4 asecs \end{tabular}			& \begin{tabular}{@{}l@{}}Semi-static; iterative\\movement of PSM and\\FSM allows translation of\\Phasecam pupils \end{tabular}\\	
\hline
\end{tabular}
\noindent
\end{center}
\end{table}

In contrast, both the LMIRcam-Fizeau\footnote{LBTI does not currently have the optics to enable LMIRcam to observe in nulling mode, so we will henceforth label the LMIRcam-Fizeau channel as simply `LMIRcam'.} and NOMIC-Fizeau channels are highly non-common from that of Phasecam (Fig. \ref{fig:nic_diag}). The LMIRcam and Phasecam channels split at a beam diverter which sits right inside the entrance of the NIC cryostat. The NOMIC-Fizeau channel splits from Phasecam two optical elements further downstream. The LMIRcam and NOMIC-Fizeau channels are blind to OPD and tip-tilt variations induced by the NIL steering mirrors and beam combiner (Table \ref{table:fiz_open_closed}). To make a Fizeau PSF with LMIRcam or NOMIC, the beams must be combined on the plane of the detector itself.

This leads to the following issues: the science and phase-sensing channels are more prone to differential OPD or tip-tilt, and the science and phase-sensing channels require \textit{separate} beam combination. Delicate alignment is required to achieve co-alignment of the coherence envelopes to image the zero-OPD interferometric PSFs on both detectors, and this alignment must be maintained for the duration of the observation.

\begin{figure}
	\centering
	\begin{subfigure}{0.4\textwidth}
		\centering
		\hspace{-1cm}
		\includegraphics[width=0.2\linewidth, trim={9.5cm, 1cm, 4.5cm, 1cm}, clip=True]{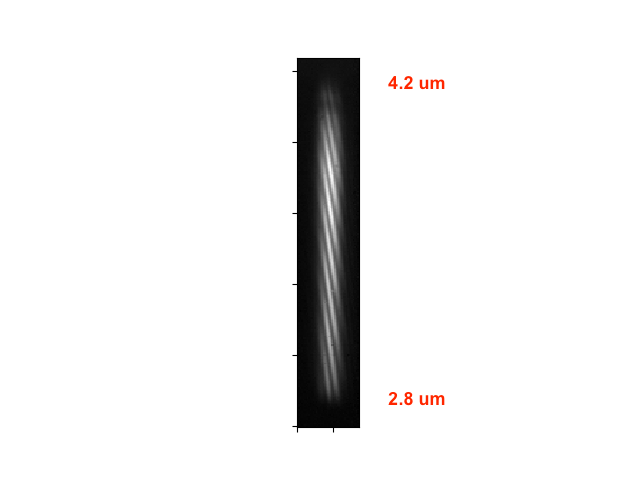}
		\hspace{-0.5cm}
		\includegraphics[width=0.2\linewidth, trim={7cm, 1cm, 7cm, 1cm}, clip=True]{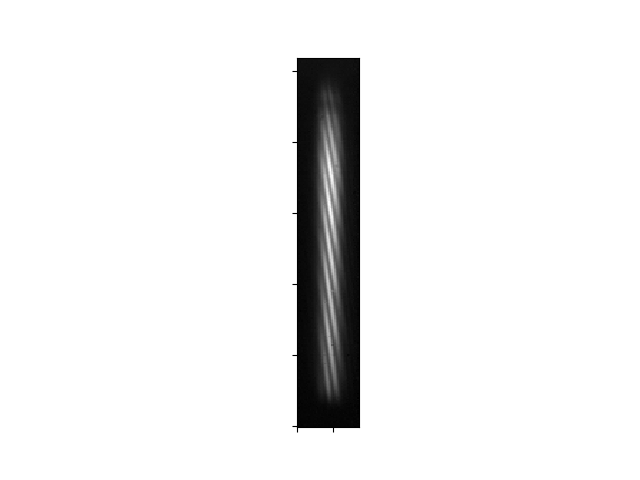}
		\includegraphics[width=0.2\linewidth, trim={7cm, 1cm, 7cm, 1cm}, clip=True]{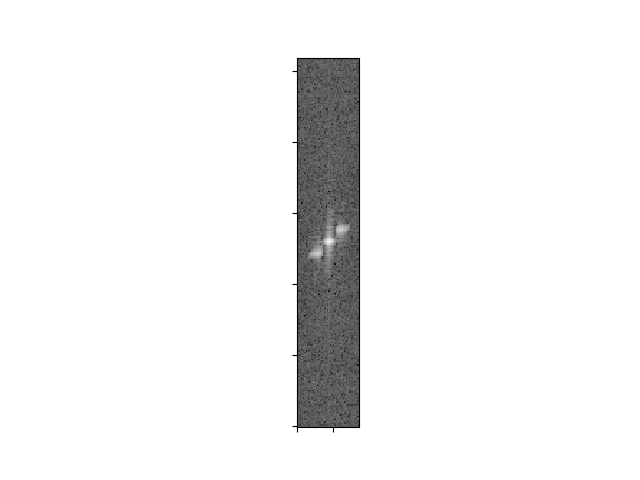}
		\includegraphics[width=0.2\linewidth, trim={7cm, 1cm, 7cm, 1cm}, clip=True]{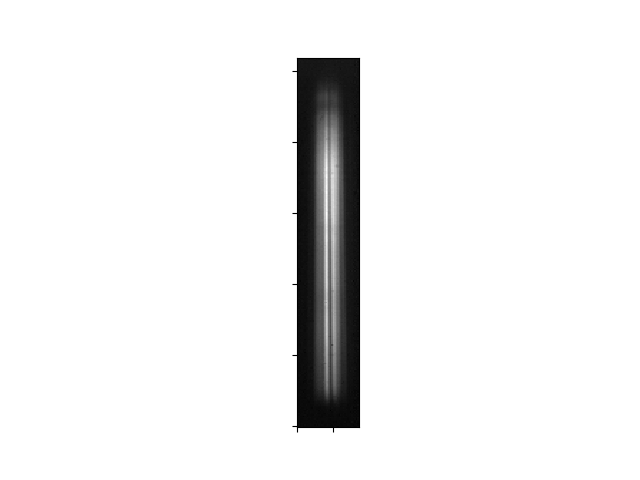}
		\includegraphics[width=0.2\linewidth, trim={7cm, 1cm, 7cm, 1cm}, clip=True]{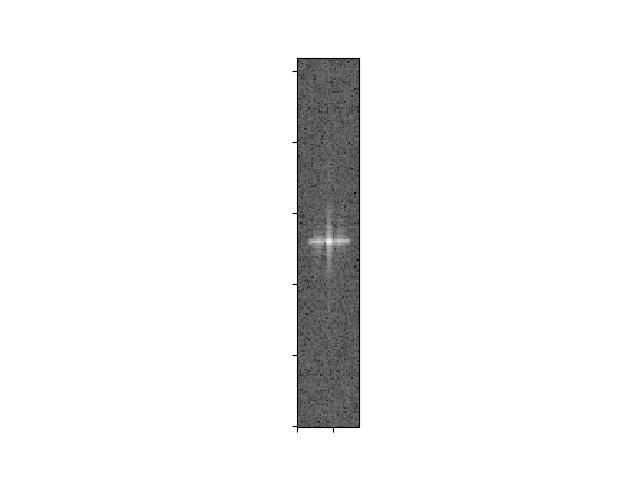}
		\caption{}
		\label{fig:barber}
	\end{subfigure}\quad
	\begin{subfigure}{0.4\textwidth}
		\centering
		\includegraphics[width=0.9\linewidth, trim={27cm, 1cm, 38.1cm, 10.5cm}, clip=True]{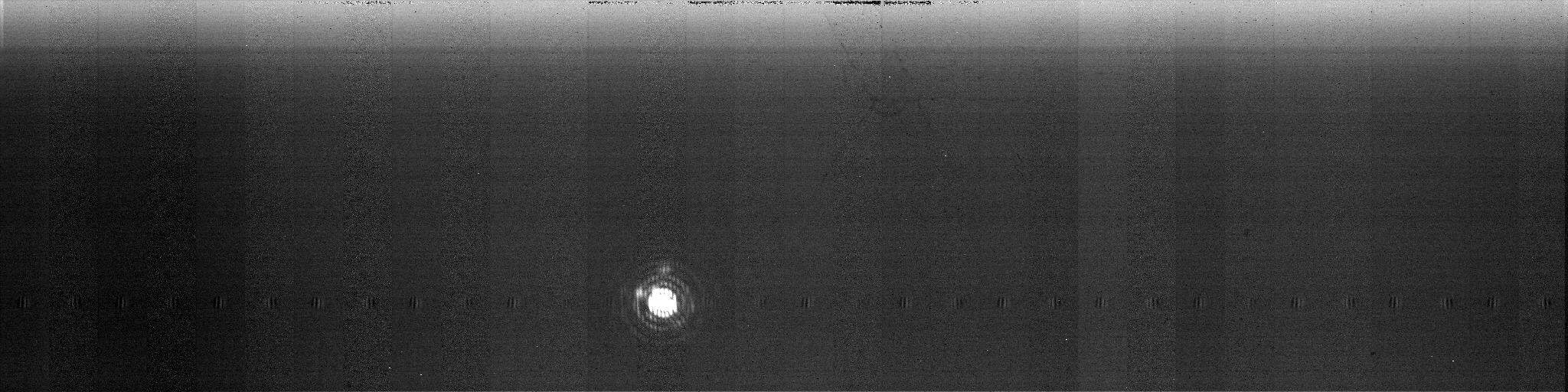}
		\caption{}
		\label{fig:phys_psf}
	\end{subfigure}\quad
\vspace{0.1cm}
\caption[]{a.) Strip-like images are, from left to right: ``barber-pole'' fringes as produced by a 2.8-4.2 $\mu$m $L$-band grism \cite{kuzmenko2012fabrication}, during the initial search for the coherence envelope on LMIRcam; a fast Fourier transform of the ``barber-pole'' fringe image, with logarithmic color scaling; an image of the fringes when they are roughly vertical after minimizing the OPD with the use of the HPC pathlength corrector mirror; and its fast Fourier transform. b.) A physical $L'$-band Fizeau PSF, complete with lumpy optical ``ghosts'', captured in May 2018. The greyscale is nonlinear to show the structure in the outer Airy rings. Bright Airy rings are  visible out to several $\lambda/D$, where $D=8.25$ m is the effective diameter of one primary mirror.}
\end{figure}

The center of the coherence envelope can be sought manually by dispersing the PSF with a grism and moving a pathlength corrector mirror (HPC) with a piezo and translation stage back and forth until obtaining ``barber pole'' fringes (which derive their shape from the wavelength-dependent phase difference). The OPD must be changed until the fringes are straightened, so as to indicate a common OPD in units of wavelength across the entire bandpass (i.e., the OPD must be zero). In the process, an algorithm may make use of fast Fourier transforms of the barber-pole images so as to sense and minimize the slant of the fringes (Fig. \ref{fig:barber}).

As a first step in controlling the PSF, Phasecam needs to be coupled to the Fizeau mode to provide a first-order correction to OPD variations. In closed loop, Phasecam sends compensating signals to the fast pathlength corrector (FPC) mirror to remove effects of instrument flexure ($<<$ Hz) and the atmosphere ($\sim$10 Hz). In order to maintain the image on Phasecam, pathlength and tip-tilt setpoints set the zero levels around which the FPC can move in piston, tip and tilt. Phasecam has limitations, however. It can only partially remove the 5-10 $\mu$m contributions of telescope resonances at 12-18 Hz and instrument (particularly UBC) vibrations at 100-150 Hz. In addition, Phasecam experiences occasional ``phase jumps'' due to the strong 2$\pi$ phase degeneracy between fringes (see Table \ref{table:fiz_open_closed} and \cite{maier2018phase} in these proceedings). Until now, those jumps have had to be counteracted manually in real-time, resulting in some loss in efficiency.  Thus Phasecam provides an important but incomplete source of control over the PSF.

To adapt Phasecam for Fizeau, additional controls are needed which will carry out the following: 1.) Telescope movements to make and maintain PSF overlap in the focal plane; 2.) Seeking and maintaining the center of the coherence envelope of the science bandpass; 3.) Calculation of OPD and tip-tilt setpoints which form the basis for the Phasecam loop which applies realtime corrective movements of mirrors for phase control; and 4.) Repeat the above steps after every telescope nod, dither, or opening of the Phasecam loop, to counteract mechanical hysteresis and optical realignments.

To be clear, at the highest level there are three control loops acting in parallel for Fizeau (or nulling) observations:

\begin{enumerate}
\item Left telescope AO control loop: wavefront correction based on secondary mirror deformation
\item Right telescope AO control loop: (the same)
\item Phasecam phase control loop: OPD and tip-tilt correction via the fast pathlength corrector mirror
\end{enumerate}

\noindent
A Fizeau calibration loop will supplement these by providing gross PSF alignment, followed by repeated coherence envelope alignment and calculation and setting of OPD and tip-tilt setpoints. 

The first control loop listed above will be upgraded in the summer of 2018 as part of the SOUL project \cite{pinna2016soul}. The second will be upgraded later on. Work is progressing to upgrade the third to supplement the $K_{S}$-band correction with $H$-band (see \cite{maier2018phase} in these proceedings). The Fizeau calibration loop does not yet exist, and the general methodology is explained in the following section. (The methodology for finding equivalent setpoints in nulling mode are described in \cite{defrere2016nulling}.)






\subsection{Envisioned Fizeau calibration loop}

The detector science readouts themselves will provide the best real-time data on OPD or tip-tilt variations that have escaped correction by Phasecam. The initialization of observations will just be an automated version of what is currently done manually: a Fizeau calibration loop should begin by moving both telescope images close together, putting in a grism, and adjusting OPD with a corrector mirror (the slow pathlength corrector in the UBC) until barber pole fringes are acquired and straightened. Once at the center of the coherence envelope, the grism is removed. The NIL beam combiner, which is upstream of Phasecam but not the LMIRcam or NOMIC Fizeau channels, is then translated to find the center of the coherence envelope on Phasecam.\footnote{It is possible to offload OPD corrections from the FPC to the NIL beam combiner, but this is inadvisable in Fizeau mode because it would separate the coherence envelopes.} Two static tuning surfaces, the Pupil and Field Steering Mirrors, can provide tip-tilt (and also OPD, if necessary) to the Phasecam pupils to optimize the fringes for tracking.

To convert information encoded in the science detector image into a form that is rapidly interpretable, we will Fourier transform the illumination on the detector. (An initial exploration of this idea was made in \cite{mccarthy2000cryogenic}.) For a perfect point source, this is essentially the same as Fourier transforming the PSF into the complex optical transfer function (OTF):

\begin{equation}
FT\left\{PSF(x,y)\right\} \equiv  OTF(\zeta,\eta) \equiv MTF(\zeta,\eta)e^{-iPTF(\zeta,\eta)}
\end{equation}

\noindent
where $(x,y)$ are coordinates in the detector plane and $(\zeta,\eta)$ are in the Fourier plane. The modulation transfer function (MTF) is the modulus of the Fourier transform, and represents the amount of retrievable contrast of the object $O$ as a function of spatial frequency (e.g., \cite{boreman2001modulation}). The phase transfer function (PTF) is the phase of the Fourier transform, and is particularly sensitive to asymmetries in the PSF. The MTF and PTF will be used to diagnose nonzero OPD, tip, tilt, and possibly other aberrations. The Fizeau calibration loop will calculate the necessary corrections and send signals to movable elements, be they mirrors or the NIL beam combiner (Table \ref{table:sources_opd_tt}).

We have made custom simulations of the LMIRcam PSF with different perturbations to tease out diagnostic quantities. Table \ref{table:fiz_img_analy} shows simulated PSFs along with their MTFs and PTFs. Some things are of note in the unabberrated Fizeau PSF in the top row. The MTF has three lobes: a central one representing the low-frequency structure of the single-mirror Airy function, and the symmetric high-frequency sidelobes which encode the narrow fringing from the separation of the two primary mirrors. The MTF lobes overlap due to the degeneracy between some baselines that can fit within a single mirror, and those across both mirrors. Indeed, the span between the mirrors was chosen to allow this continuous sampling in the first place (see Sec. \ref{sec:intro}). The PTF is just a flat plane where the OTF is nonzero, because the PSF is symmetric.

When the center of the coherence envelope is on either LMIRcam or NOMIC, the illumination on the other science detector should also be close to zero OPD, because both the LMIRcam and NOMIC optical trains were built to have the same focus at the detector. There may, however, be some residual defocus between them. There are currently no optics for changing the OPD between LMIRcam and NOMIC, so if Fizeau on \textit{both} detectors is desired, then a Fizeau calibration loop using the readouts of only one detector could  have the option of choosing the detector of the highest science priority and allow the other detector to piggyback, or choose the detector with the shortest integration times (for faster correction) or greatest Fizeau calibration loop stability (for continuous correction) and impose an OPD offset to accommodate the other detector. 


\begin{figure}
\begin{center}
\includegraphics[width=0.8\linewidth]{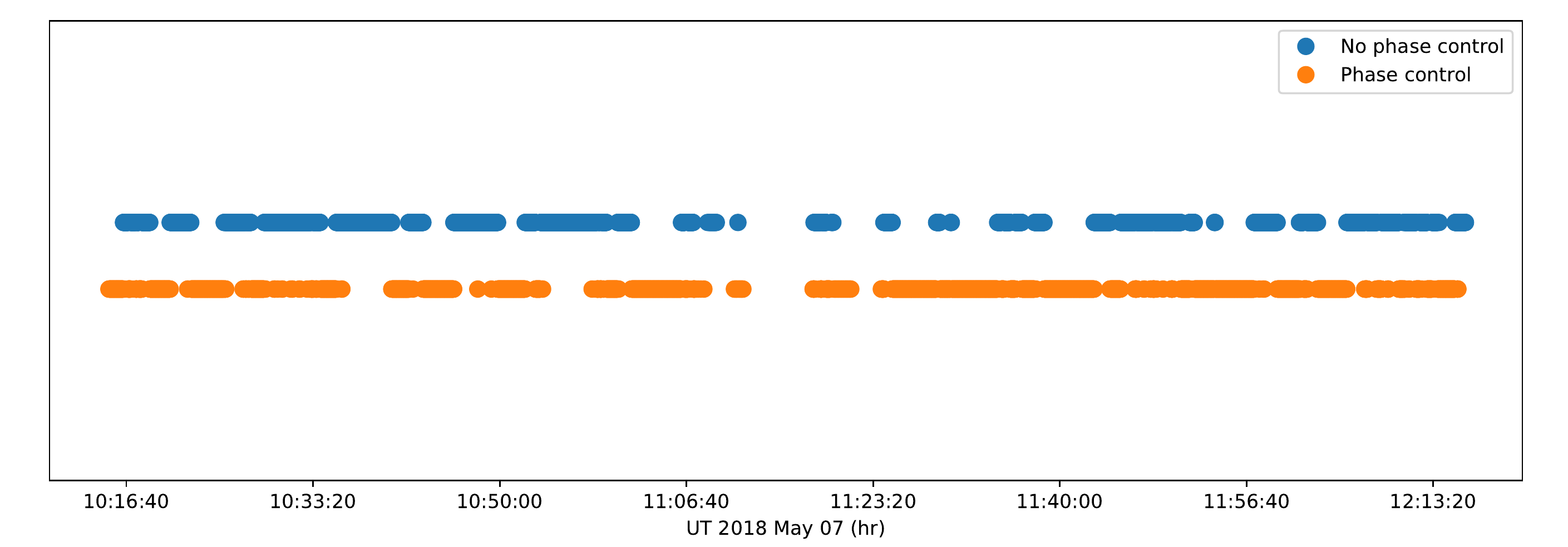}
\caption[]{A boolean plot showing when Fizeau science frames were taken with the Phasecam loop closed or open during a science observation in May 2018. Phasecam stayed closed continuously for $\sim$few minutes at a time.}
\label{fig:pcclosed_bool}
\end{center}
\end{figure}

\begin{table}
\begin{center}
\caption{Qualitative comparison of physical, on-sky PSFs when Phasecam is in open or closed loop} 
\label{table:fiz_open_closed}
\begin{tabular}{| c | c | c | c | c | c | c |}
\hline
\rotatebox[origin=l]{90}{
    Open loop
} &
\includegraphics[trim={4cm, 2cm, 3.5cm, 1.5cm}, clip=True, width=0.13\linewidth]{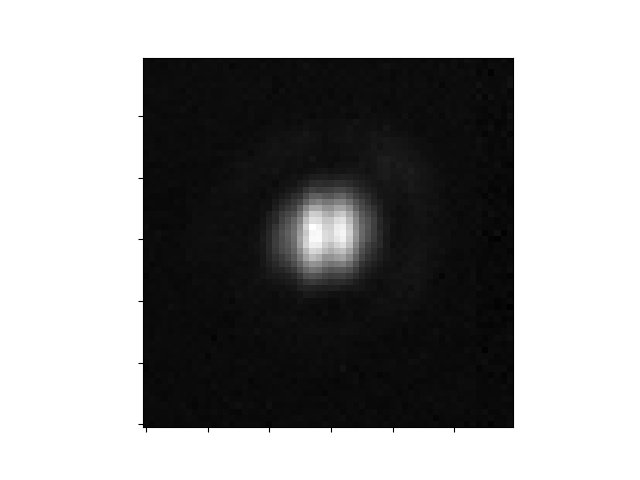} & 
\includegraphics[trim={4cm, 2cm, 3.5cm, 1.5cm}, clip=True, width=0.13\linewidth]{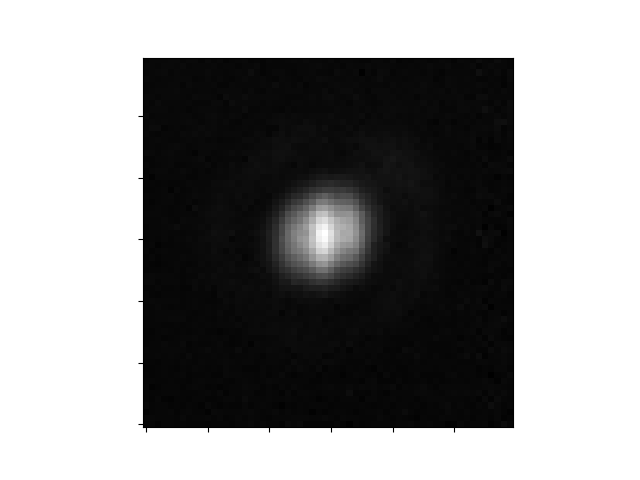} & 
\includegraphics[trim={4cm, 2cm, 3.5cm, 1.5cm}, clip=True, width=0.13\linewidth]{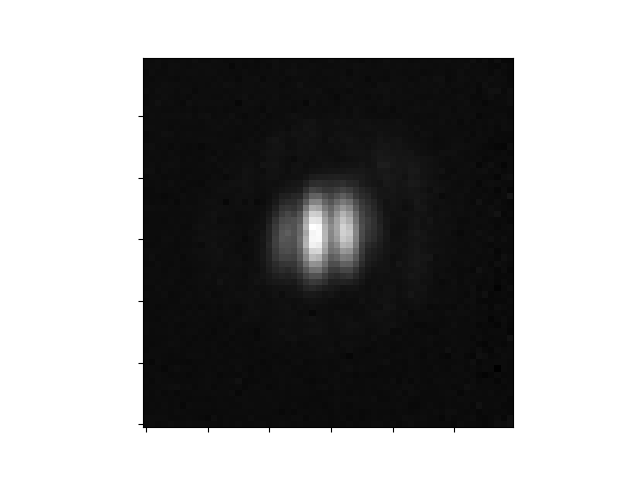} & 
\includegraphics[trim={4cm, 2cm, 3.5cm, 1.5cm}, clip=True, width=0.13\linewidth]{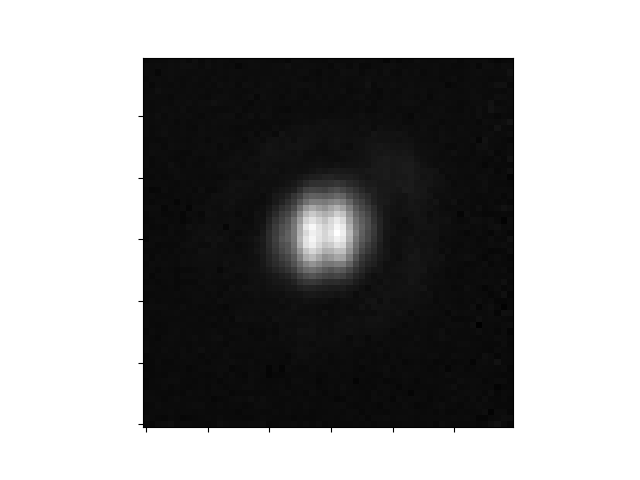} & 
\includegraphics[trim={4cm, 2cm, 3.5cm, 1.5cm}, clip=True, width=0.13\linewidth]{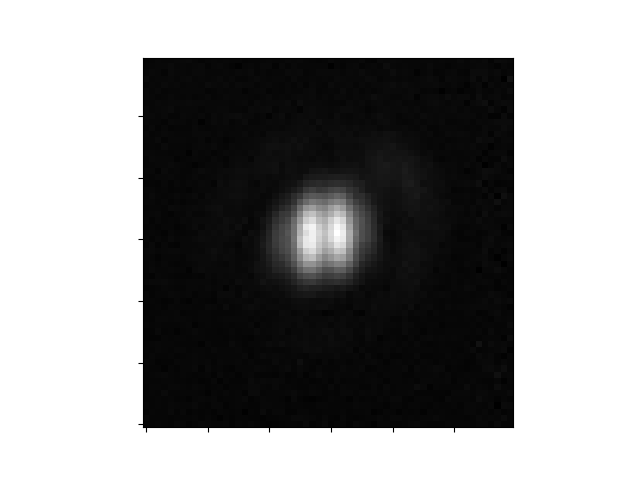} & 
\includegraphics[trim={4cm, 2cm, 3.5cm, 1.5cm}, clip=True, width=0.13\linewidth]{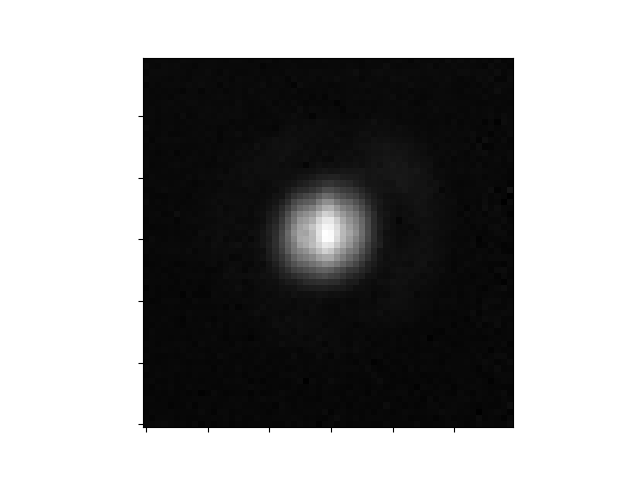} \\
\hline

\rotatebox[origin=l]{90}{
    Closed loop
} &
\includegraphics[trim={4cm, 2cm, 3.5cm, 1.5cm}, clip=True, width=0.13\linewidth]{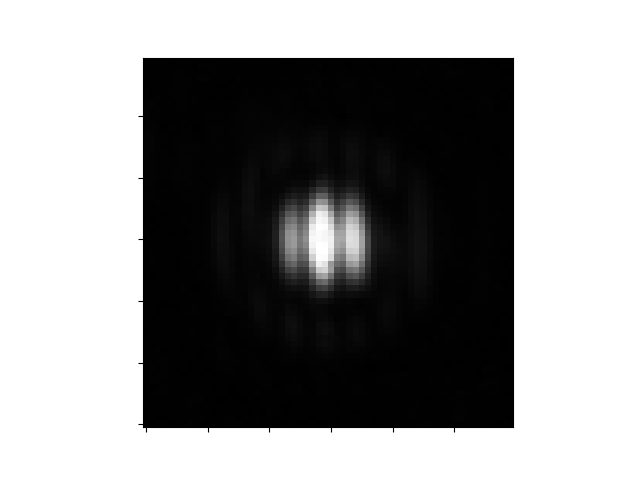} & 
\includegraphics[trim={4cm, 2cm, 3.5cm, 1.5cm}, clip=True, width=0.13\linewidth]{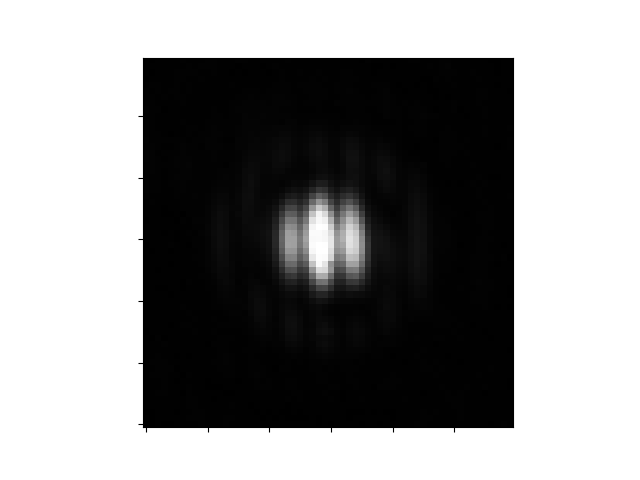} & 
\includegraphics[trim={4cm, 2cm, 3.5cm, 1.5cm}, clip=True, width=0.13\linewidth]{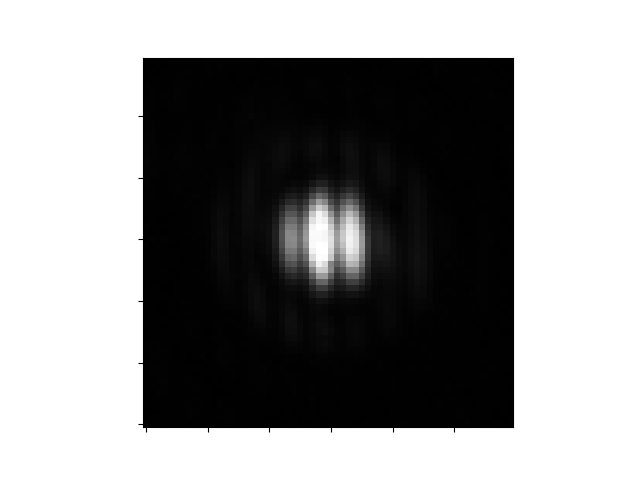} & 
\includegraphics[trim={4cm, 2cm, 3.5cm, 1.5cm}, clip=True, width=0.13\linewidth]{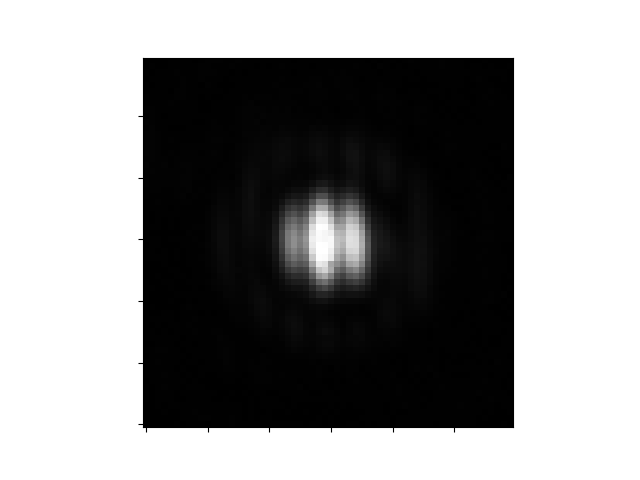} & 
\includegraphics[trim={4cm, 2cm, 3.5cm, 1.5cm}, clip=True, width=0.13\linewidth]{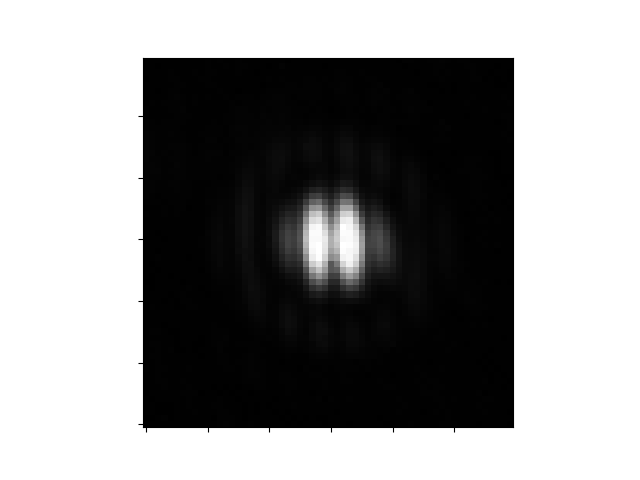} & 
\includegraphics[trim={4cm, 2cm, 3.5cm, 1.5cm}, clip=True, width=0.13\linewidth]{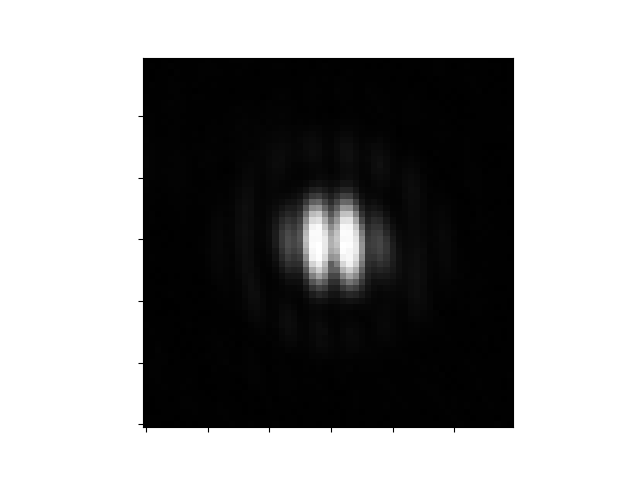} \\
\hline
& $t = 0.00$ sec & $t=0.90$ sec & $t=1.81$ sec & $t=2.71$ sec & $t=3.61$ sec & $t=4.51$ sec  \\
\hline
\end{tabular}
\end{center}
\caption*{\small{Physical PSFs obtained on-sky in May 2018, during the first sustained phase-controlled Fizeau observations with LBTI. Note the phase jump between $t=2.71$ sec and $t=3.61$ sec in the bottom row.}} \label{tab:sometab}
\noindent
\end{table}


\begin{table}
\begin{center}
\caption{Simulated Fizeau image diagnostics} 
\label{table:fiz_img_analy}
\begin{tabular}{| m{3.2cm} | m{2.1cm} | m{2.1cm} | m{2.1cm} | m{5.3cm} |}
\hline
& PSF & MTF & PTF & Potential diagnostic \\
\hline
`Perfect' Fizeau \rule{0pt}{65pt} & \includegraphics[trim={3.7cm, 1.5cm, 3cm, 1.5cm}, clip=False, width=20mm, height=20mm]{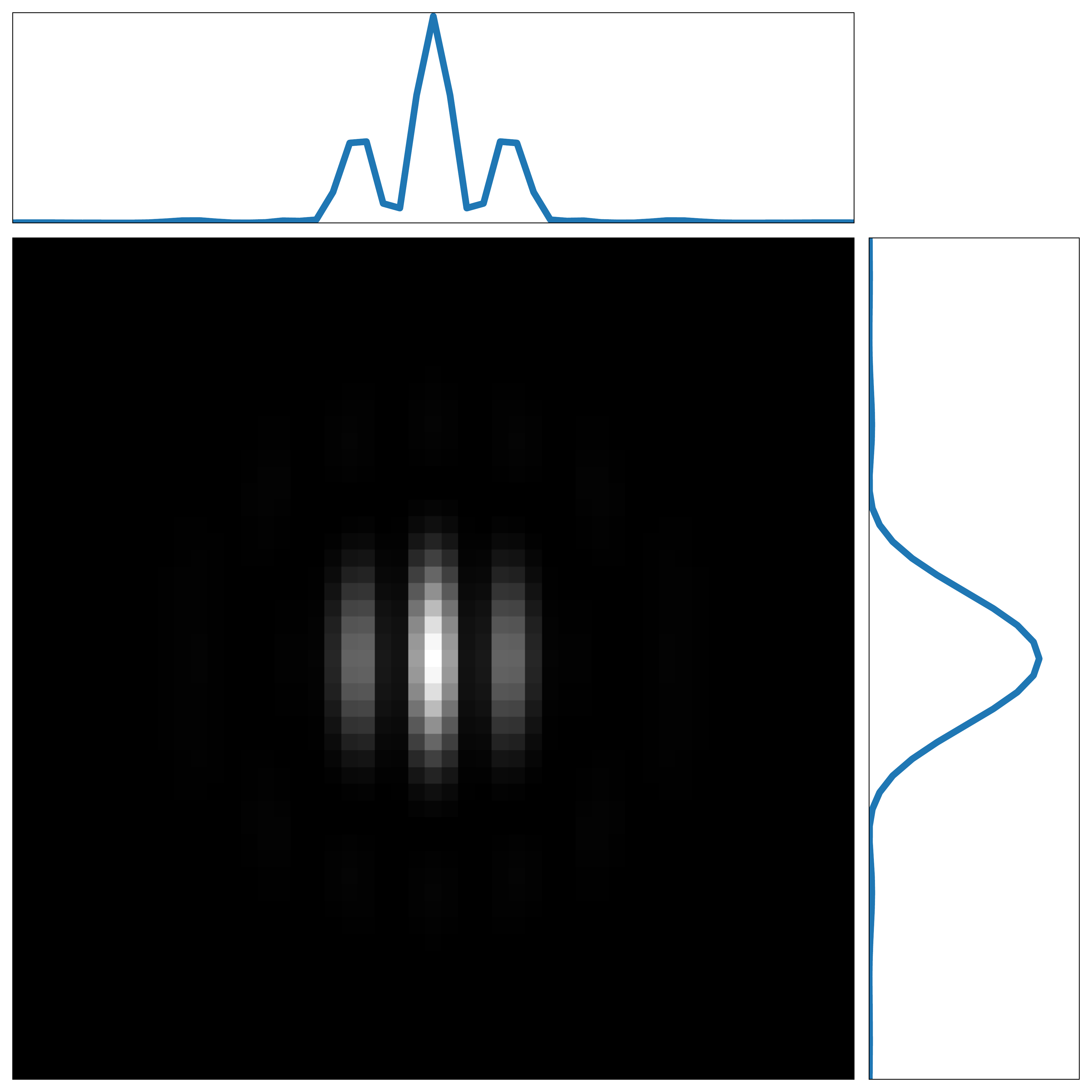} & \includegraphics[trim={3.7cm, 1.5cm, 3cm, 1.5cm}, clip=False, width=20mm, height=20mm]{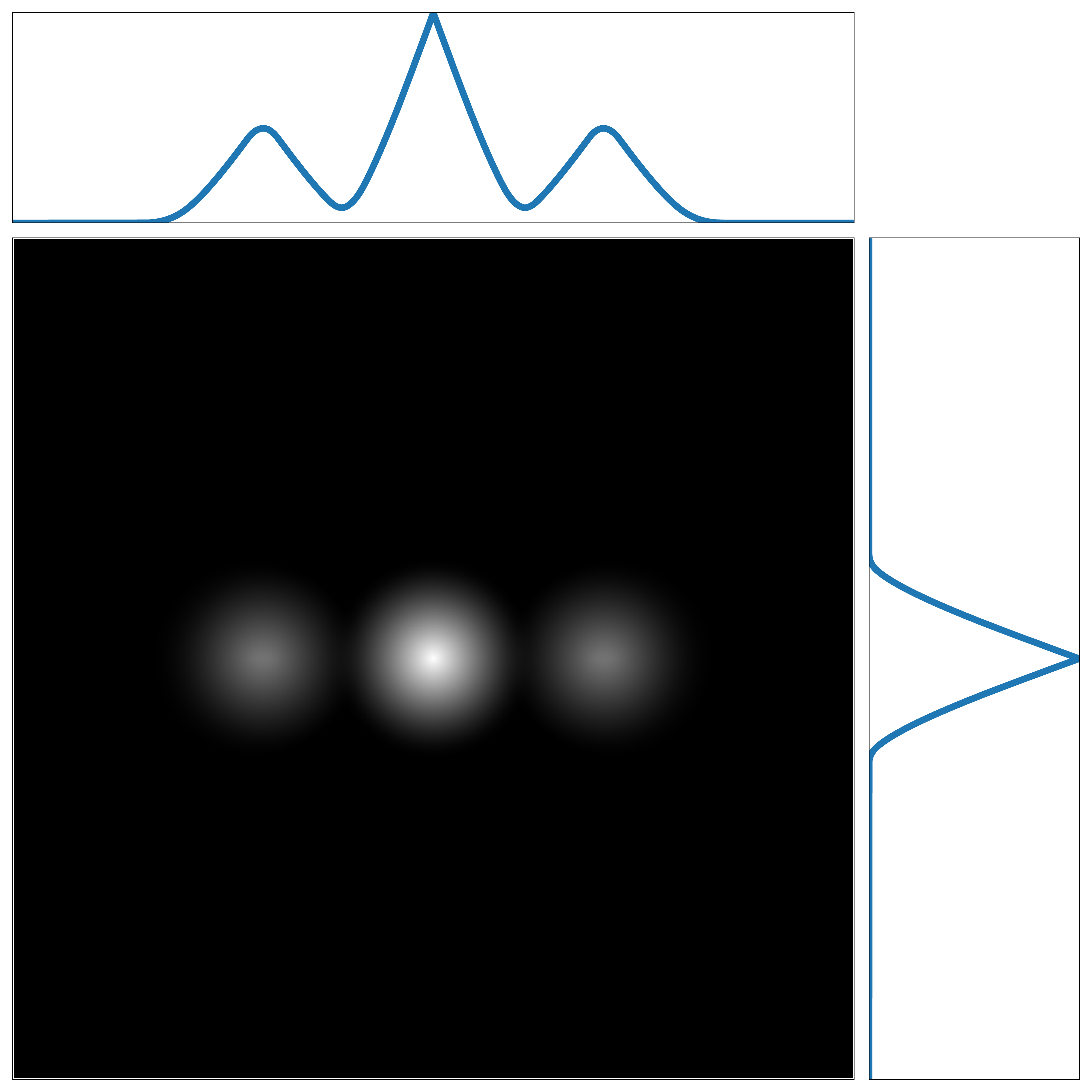} & \includegraphics[trim={3.7cm, 1.5cm, 3cm, 1.5cm}, clip=False, width=20mm, height=20mm]{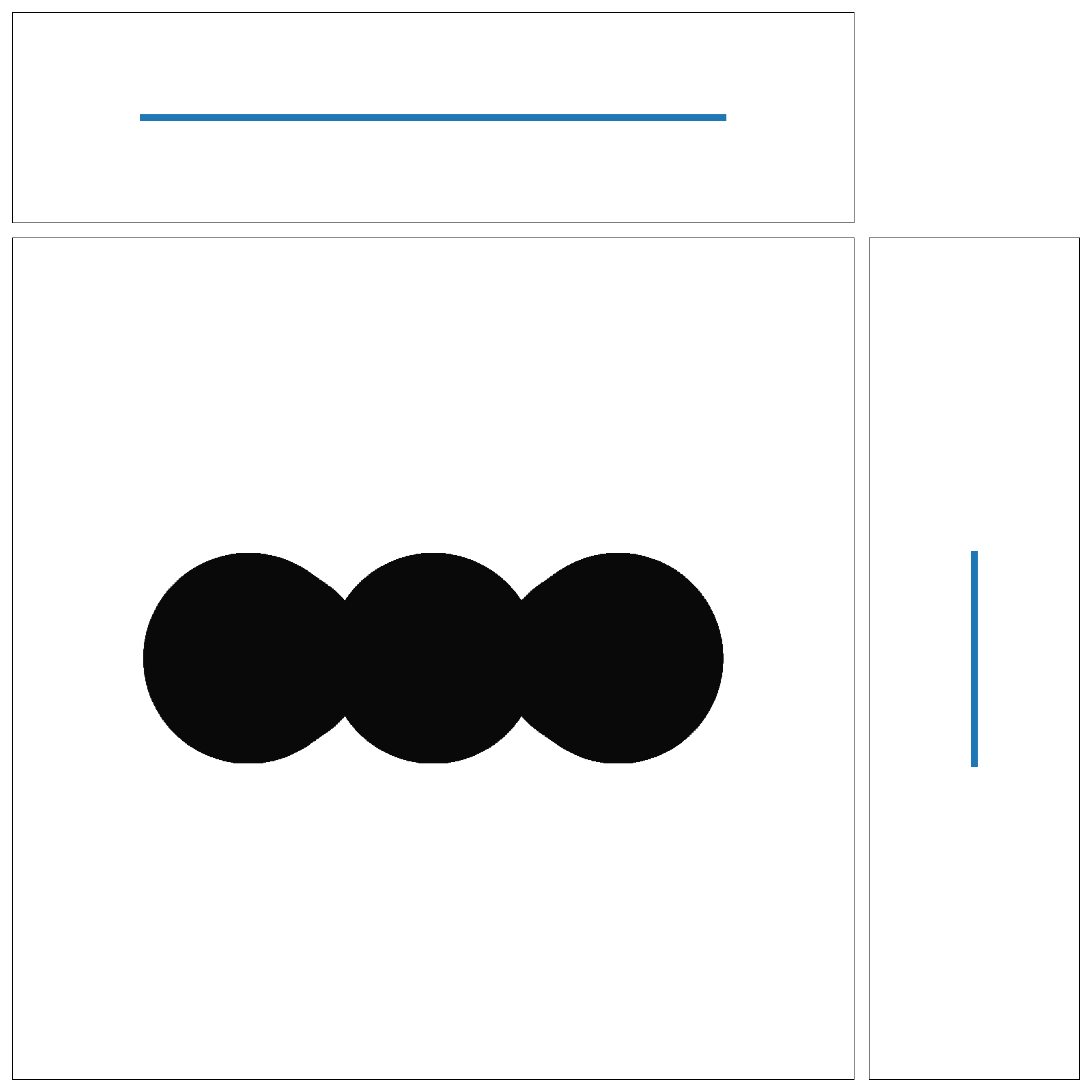} & --\\
\hline
\rule{0pt}{65pt} Diff. tip ($\pm$y), 0.03''  & \includegraphics[trim={3.7cm, 1.5cm, 3cm, 1.5cm}, clip=False, width=20mm, height=20mm]{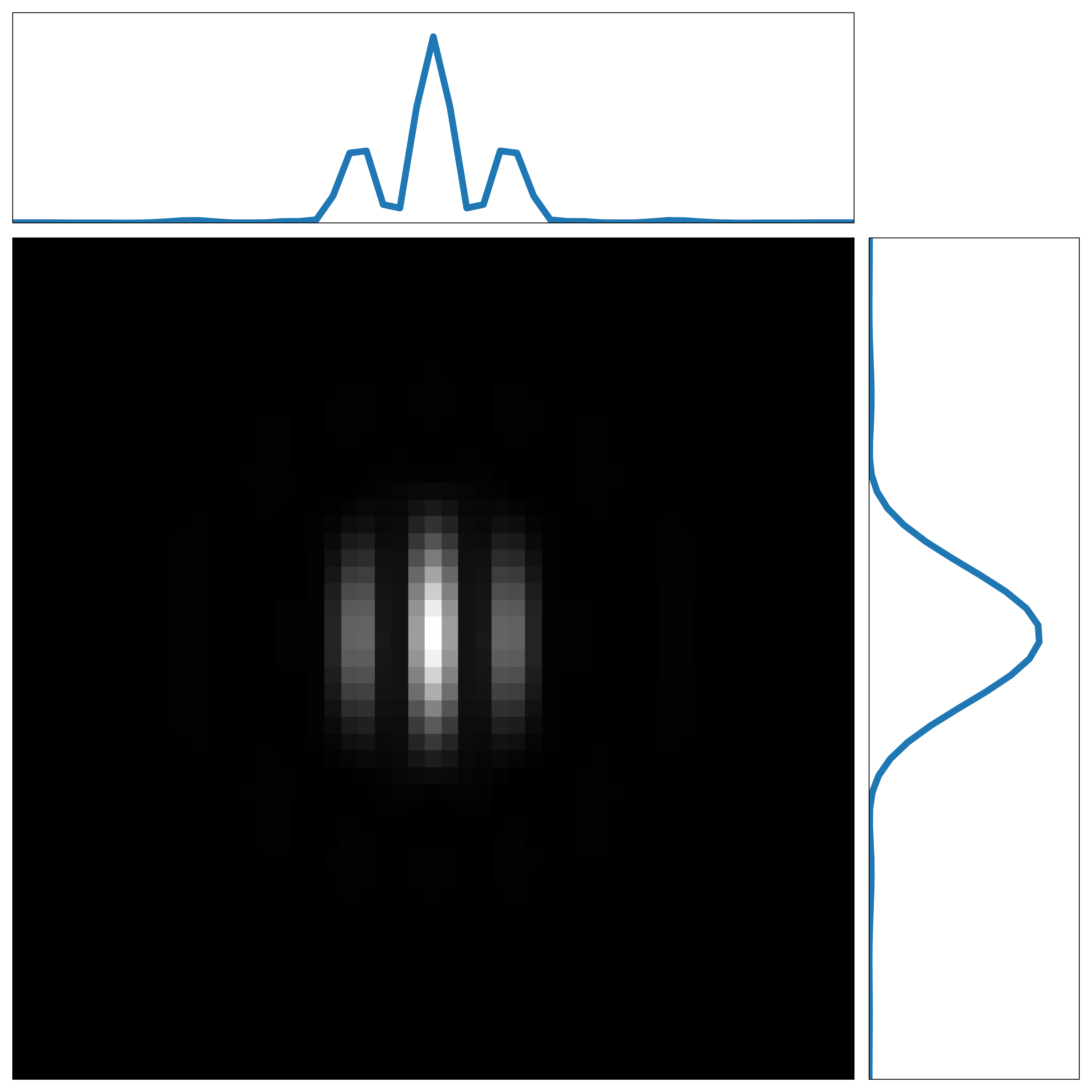} & \includegraphics[trim={3.7cm, 1.5cm, 3cm, 1.5cm}, clip=False, width=20mm, height=20mm]{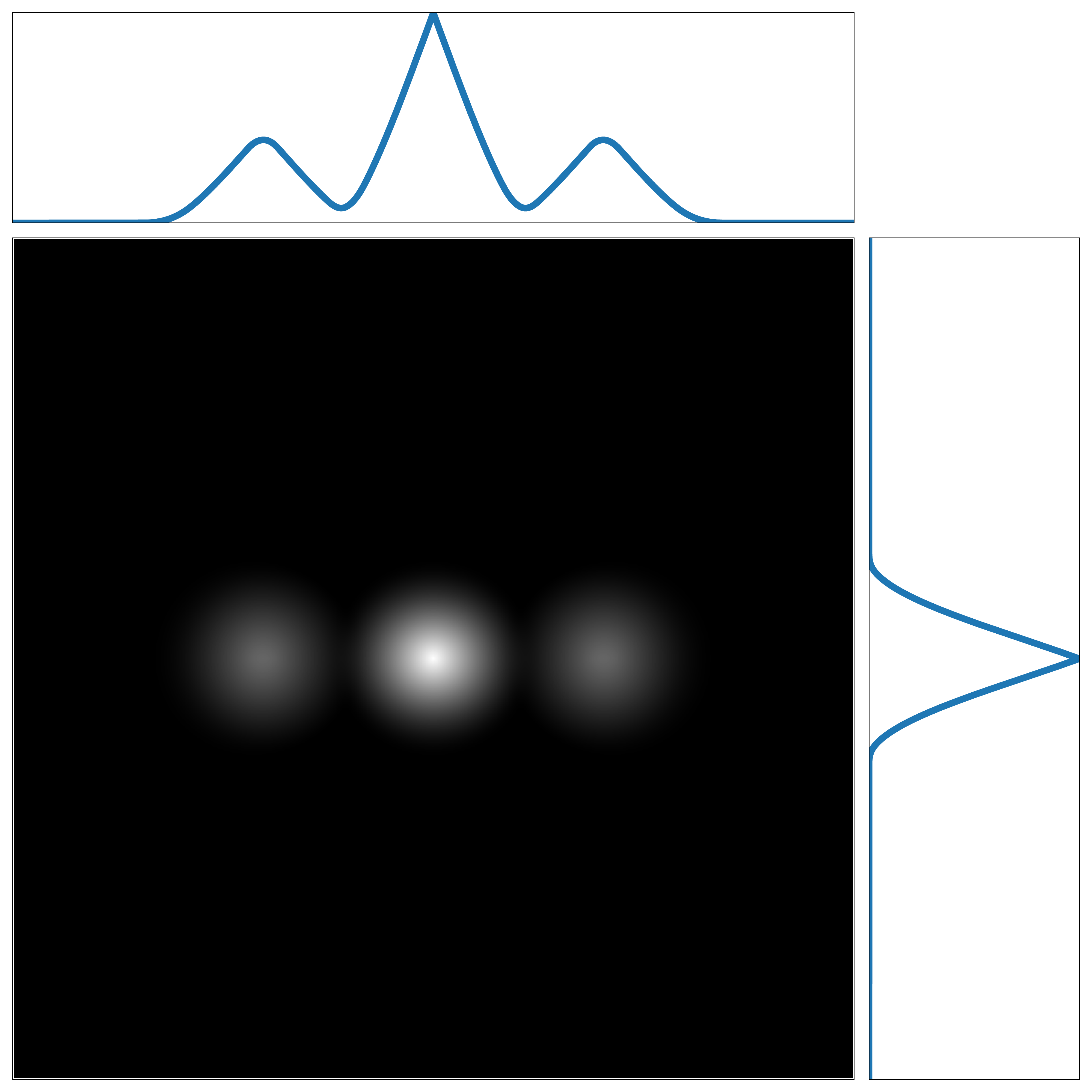} & \includegraphics[trim={3.7cm, 1.5cm, 3cm, 1.5cm}, clip=False, width=20mm, height=20mm]{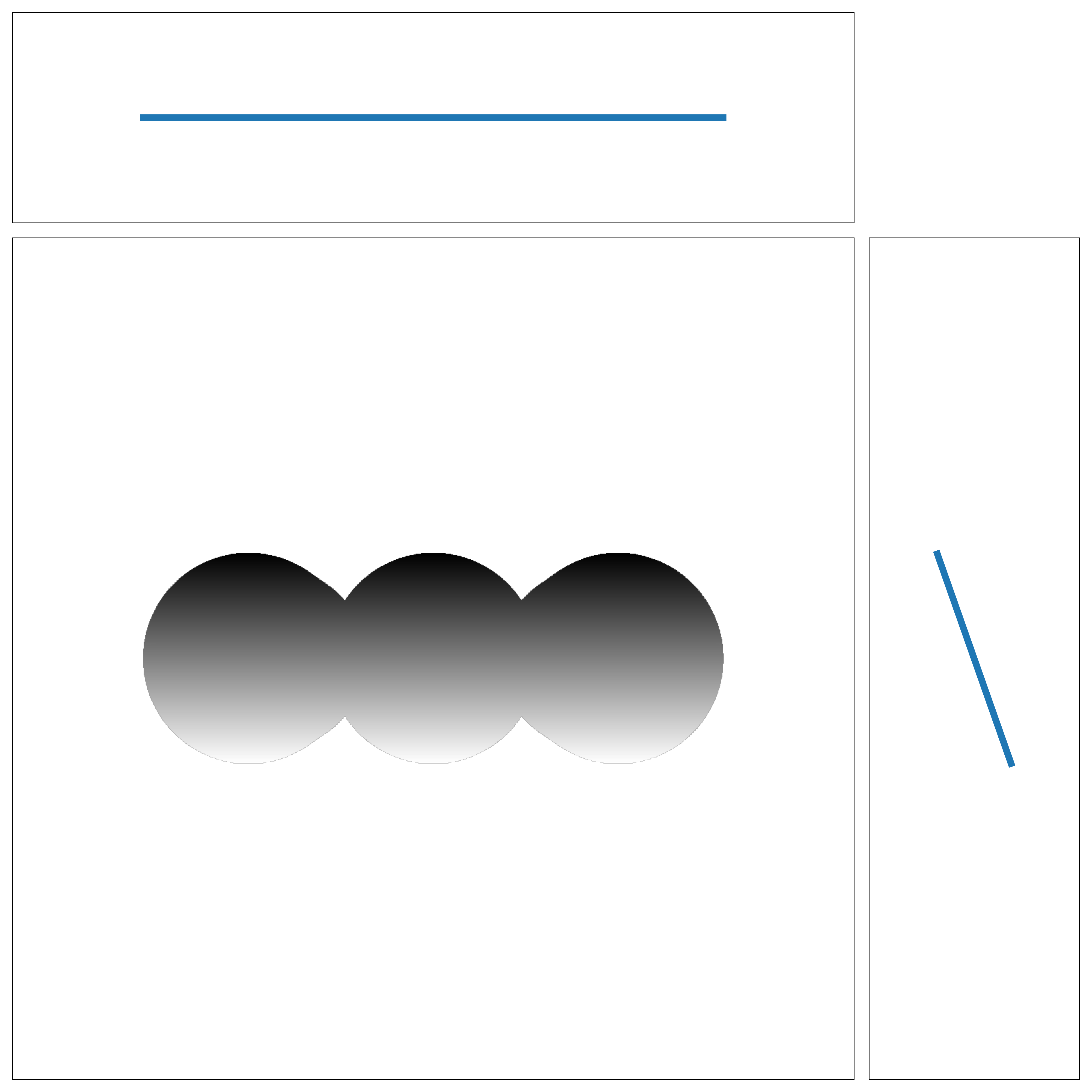} & Global, continuous phase gradient in y across all nodes \\
\hline
\rule{0pt}{65pt} Diff. tilt ($\pm$x), 0.03''  & \includegraphics[trim={3.7cm, 1.5cm, 3cm, 1.5cm}, clip=False, width=20mm, height=20mm]{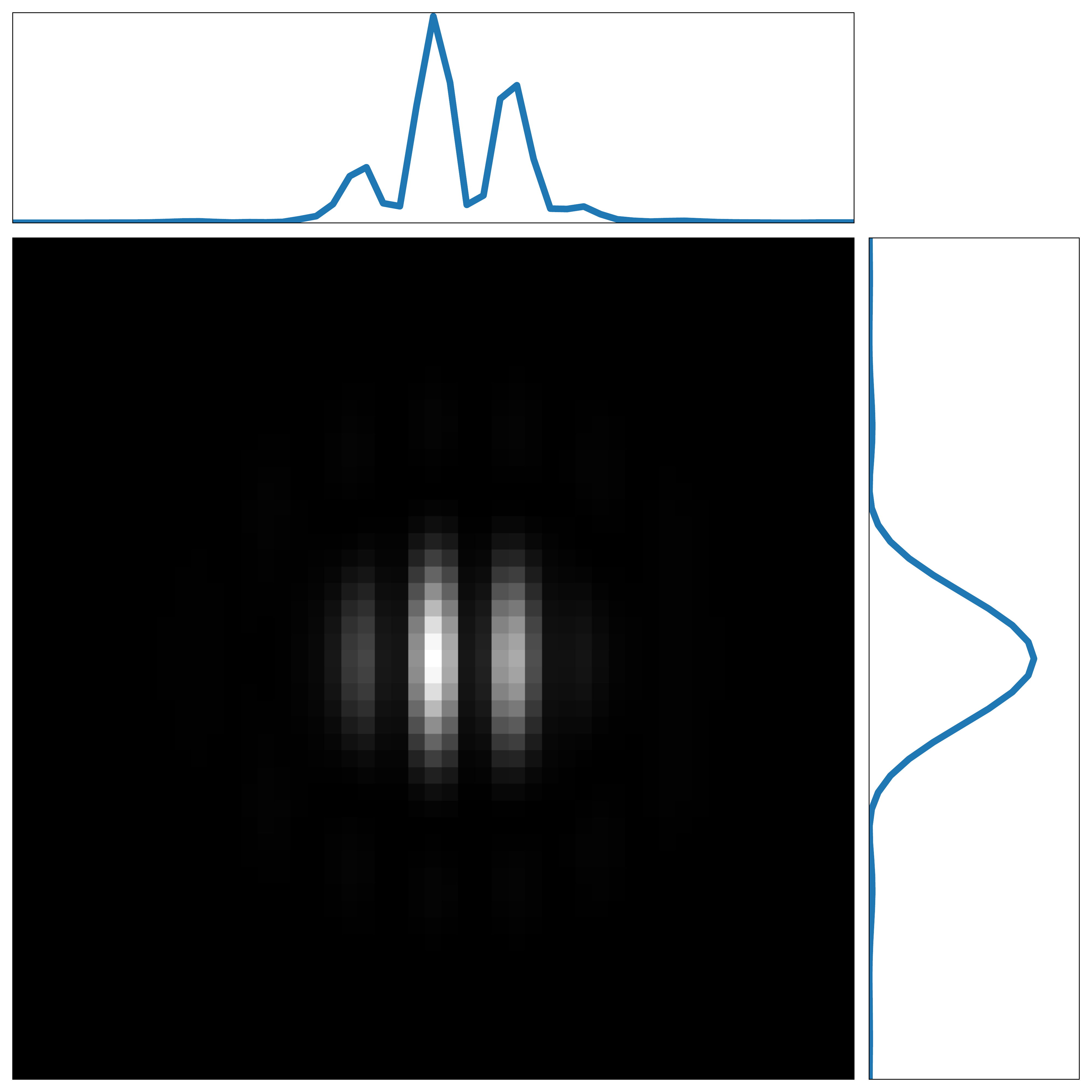} & \includegraphics[trim={3.7cm, 1.5cm, 3cm, 1.5cm}, clip=False, width=20mm, height=20mm]{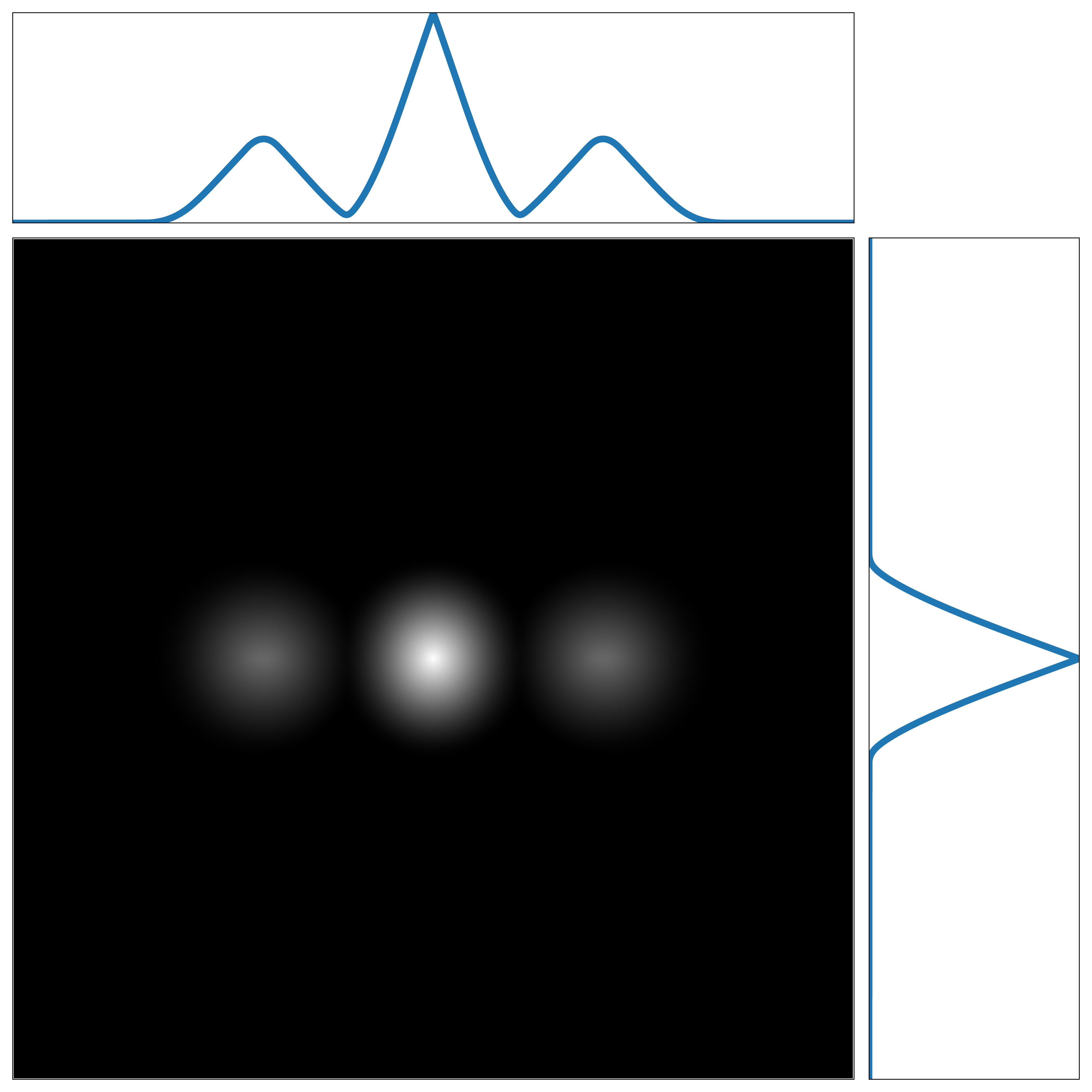} & \includegraphics[trim={3.7cm, 1.5cm, 3cm, 1.5cm}, clip=False, width=20mm, height=20mm]{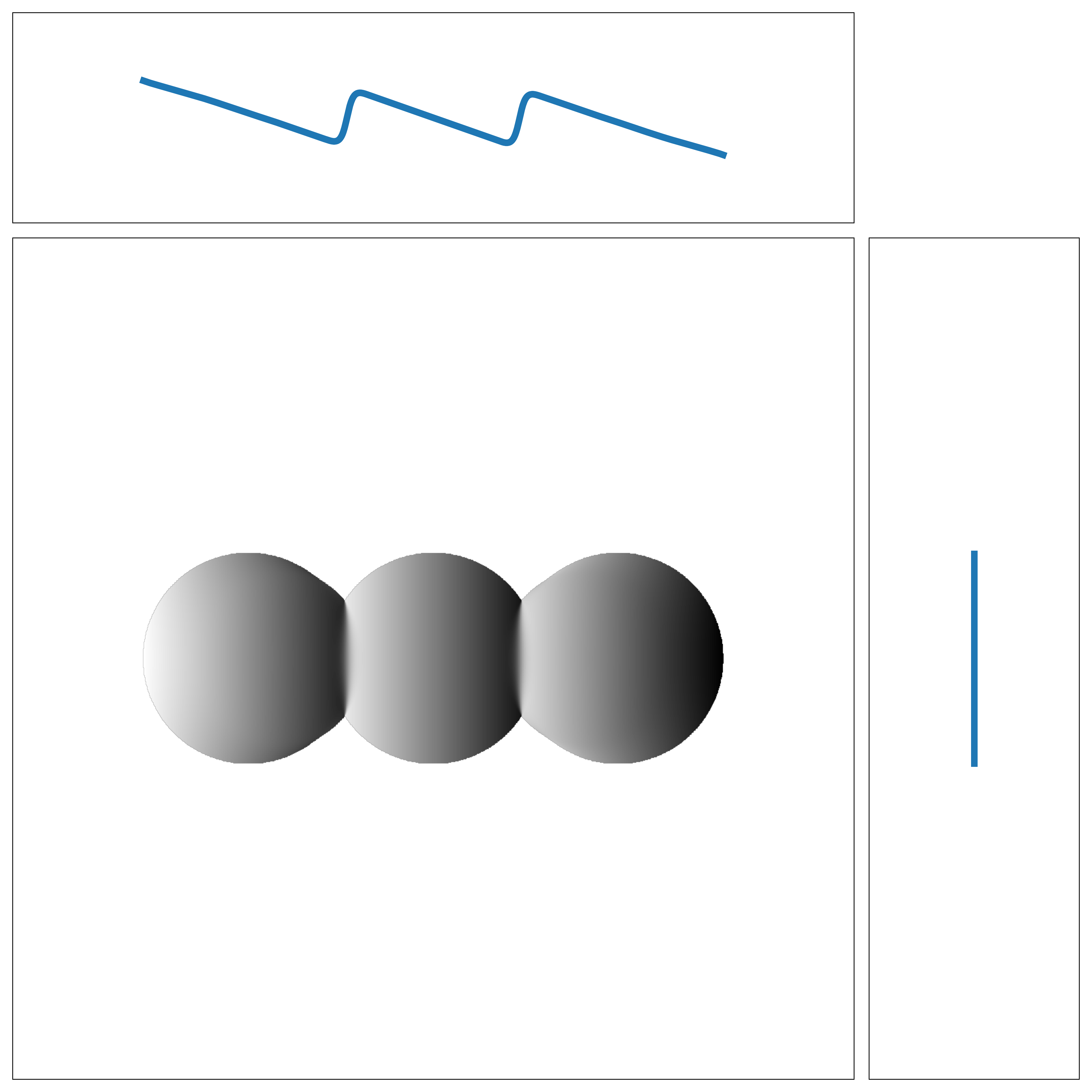} & Phase gradient in x across individual nodes\\
\hline
\rule{0pt}{65pt} Small OPD, 0.5 $\mu$m  & \includegraphics[trim={3.7cm, 1.5cm, 3cm, 1.5cm}, clip=False, width=20mm, height=20mm]{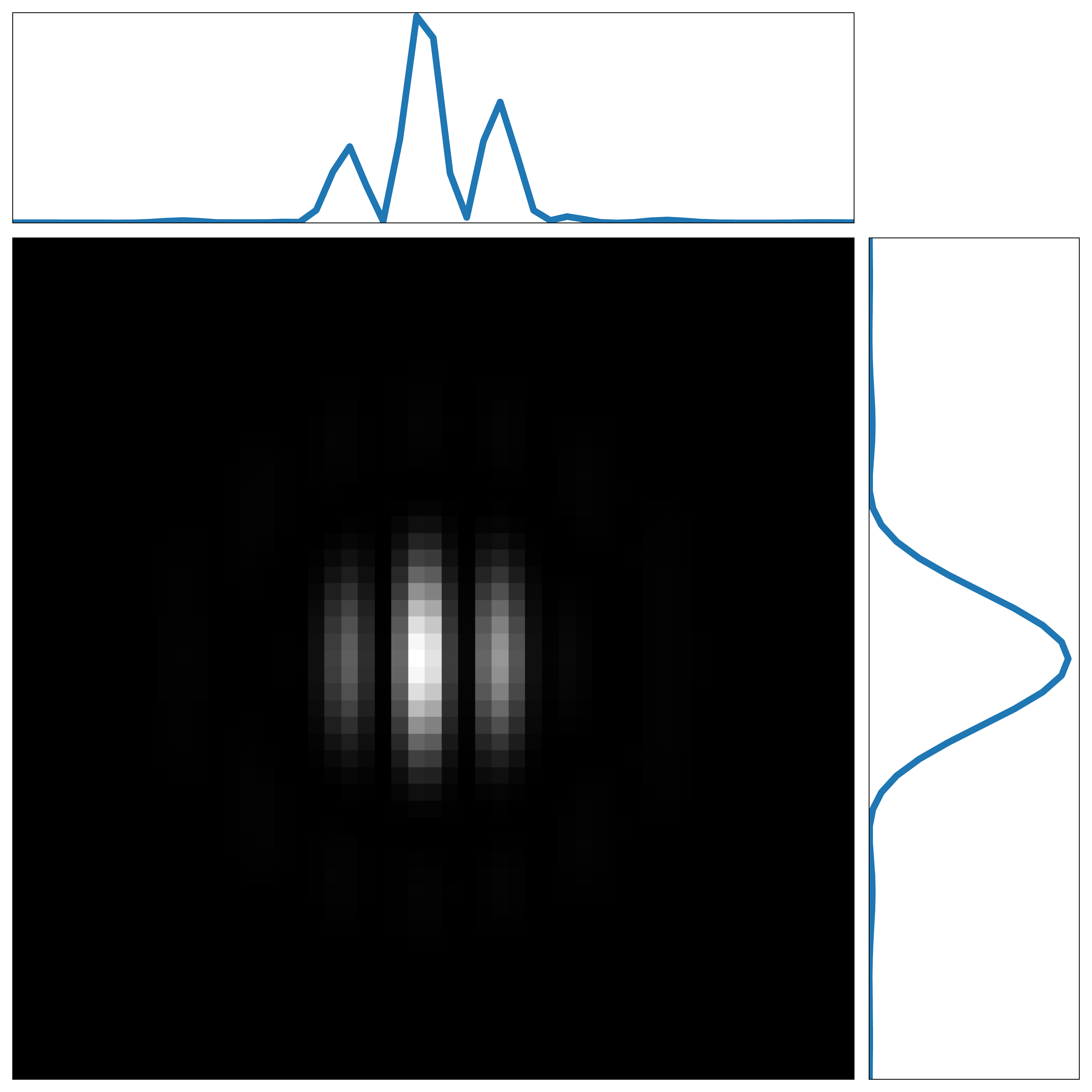} & \includegraphics[trim={3.7cm, 1.5cm, 3cm, 1.5cm}, clip=False, width=20mm, height=20mm]{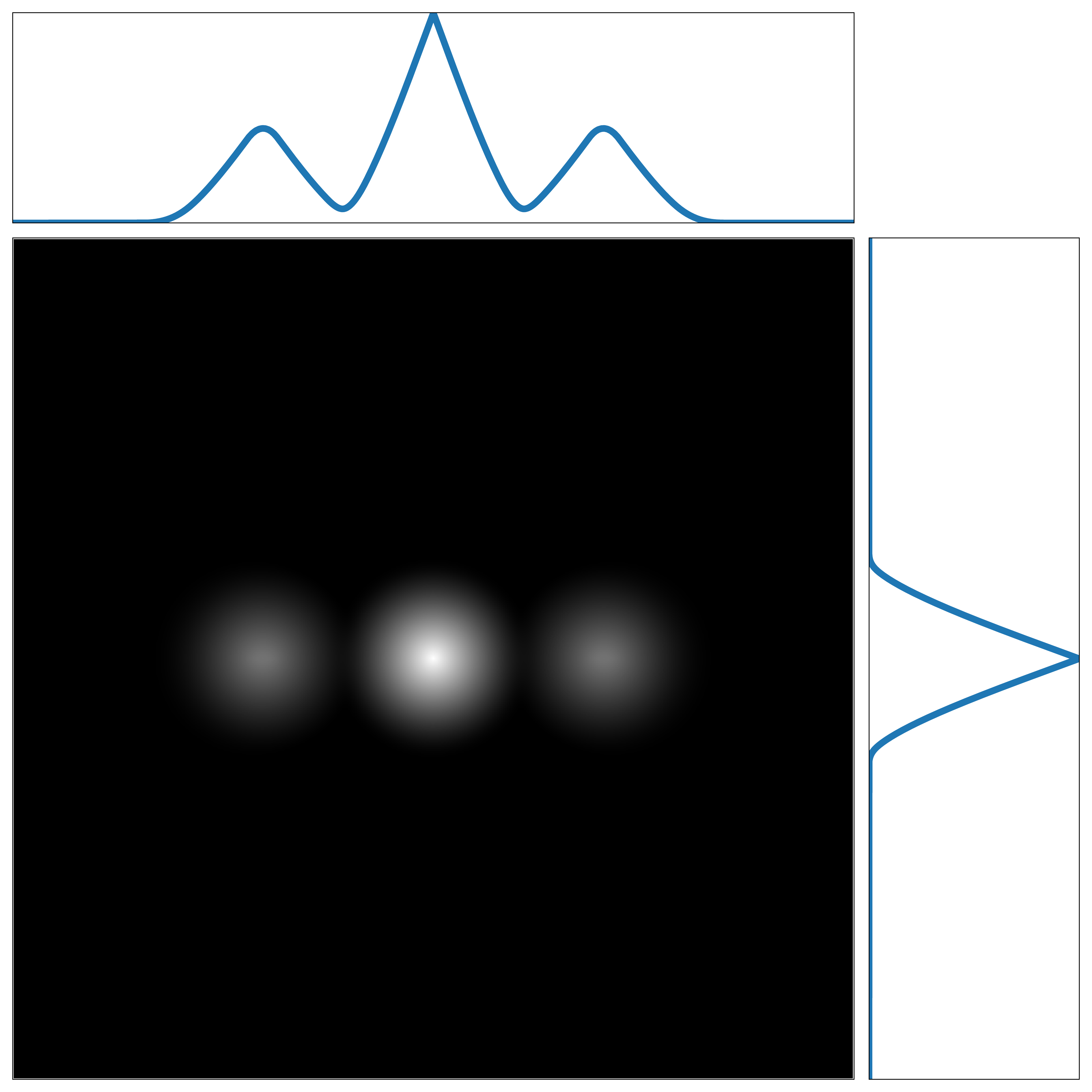} & \includegraphics[trim={3.7cm, 1.5cm, 3cm, 1.5cm}, clip=False, width=20mm, height=20mm]{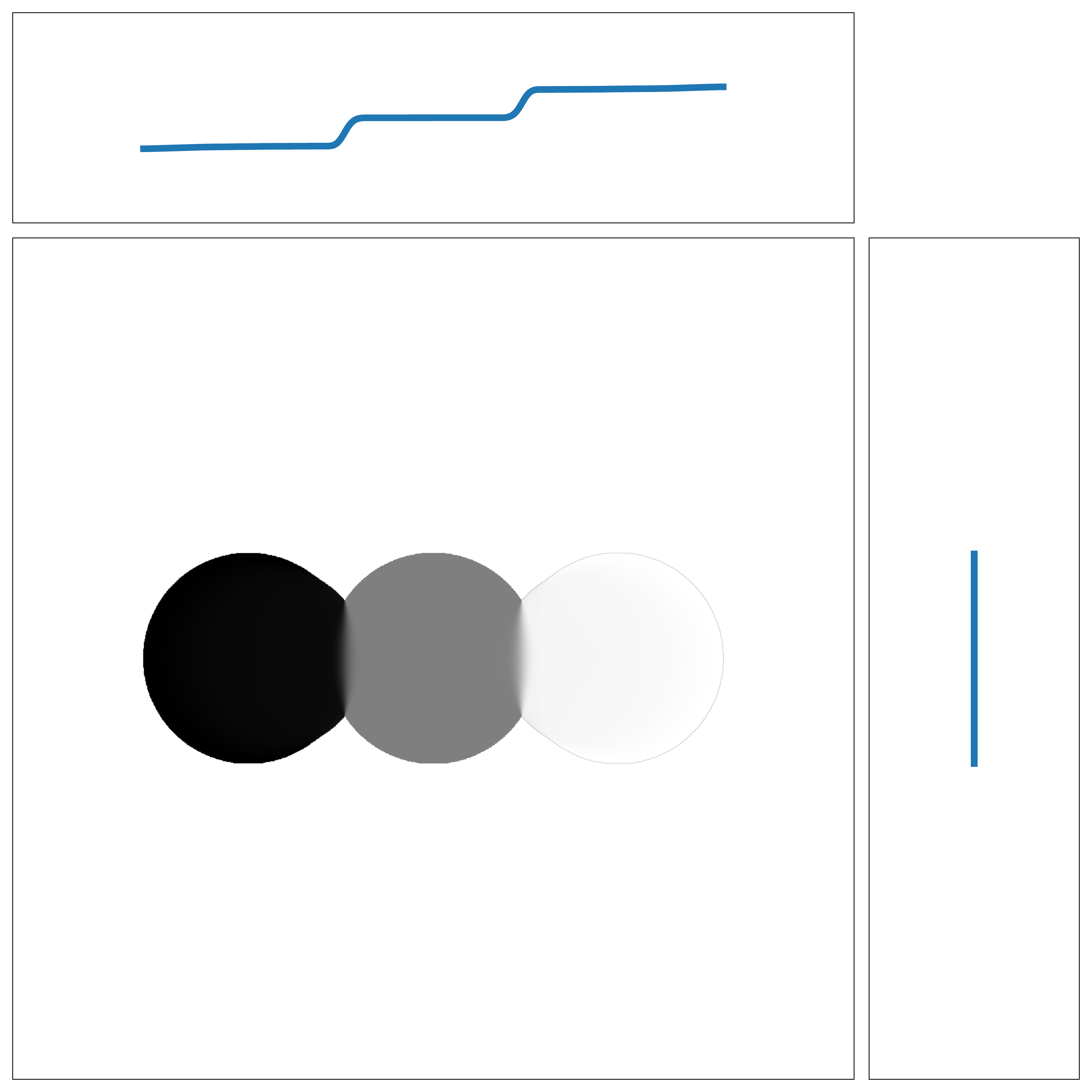} & Global phase gradient, but with flat phase value in each node \\
\hline
\rule{0pt}{65pt} Large OPD, 50 $\mu$m & \includegraphics[trim={3.7cm, 1.5cm, 3cm, 1.5cm}, clip=False, width=20mm, height=20mm]{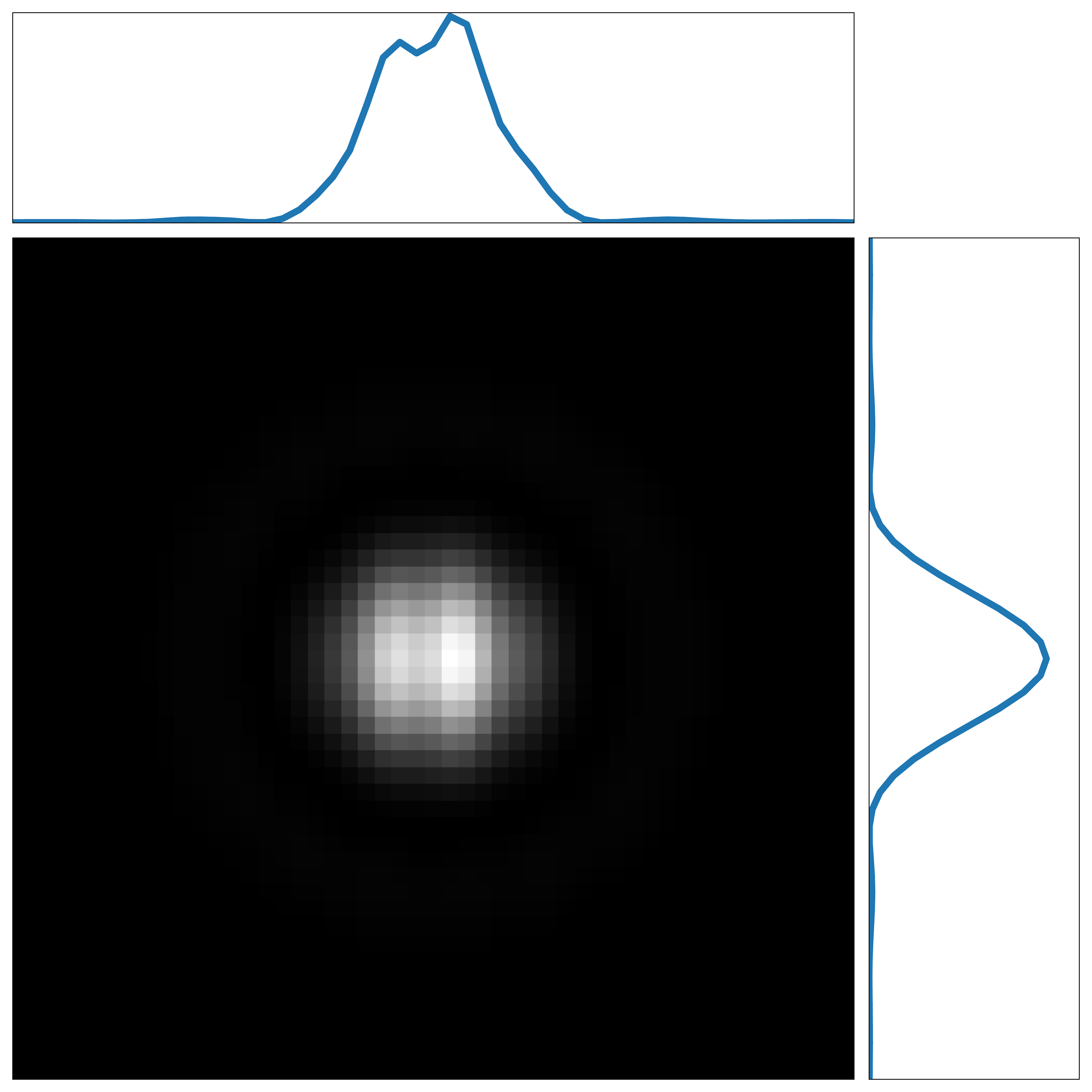} & \includegraphics[trim={3.7cm, 1.5cm, 3cm, 1.5cm}, clip=False, width=20mm, height=20mm]{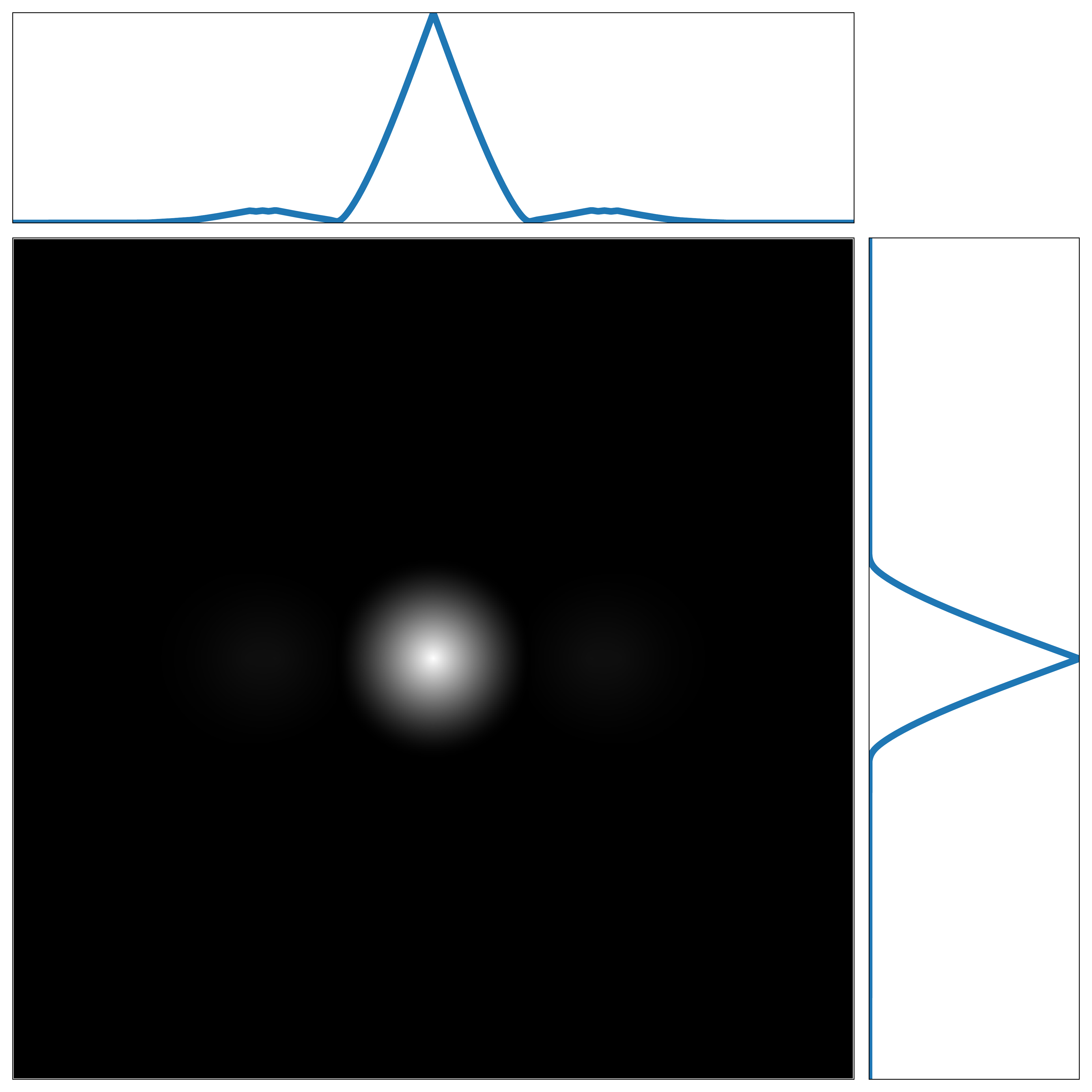} & \includegraphics[trim={3.7cm, 1.5cm, 3cm, 1.5cm}, clip=False, width=20mm, height=20mm]{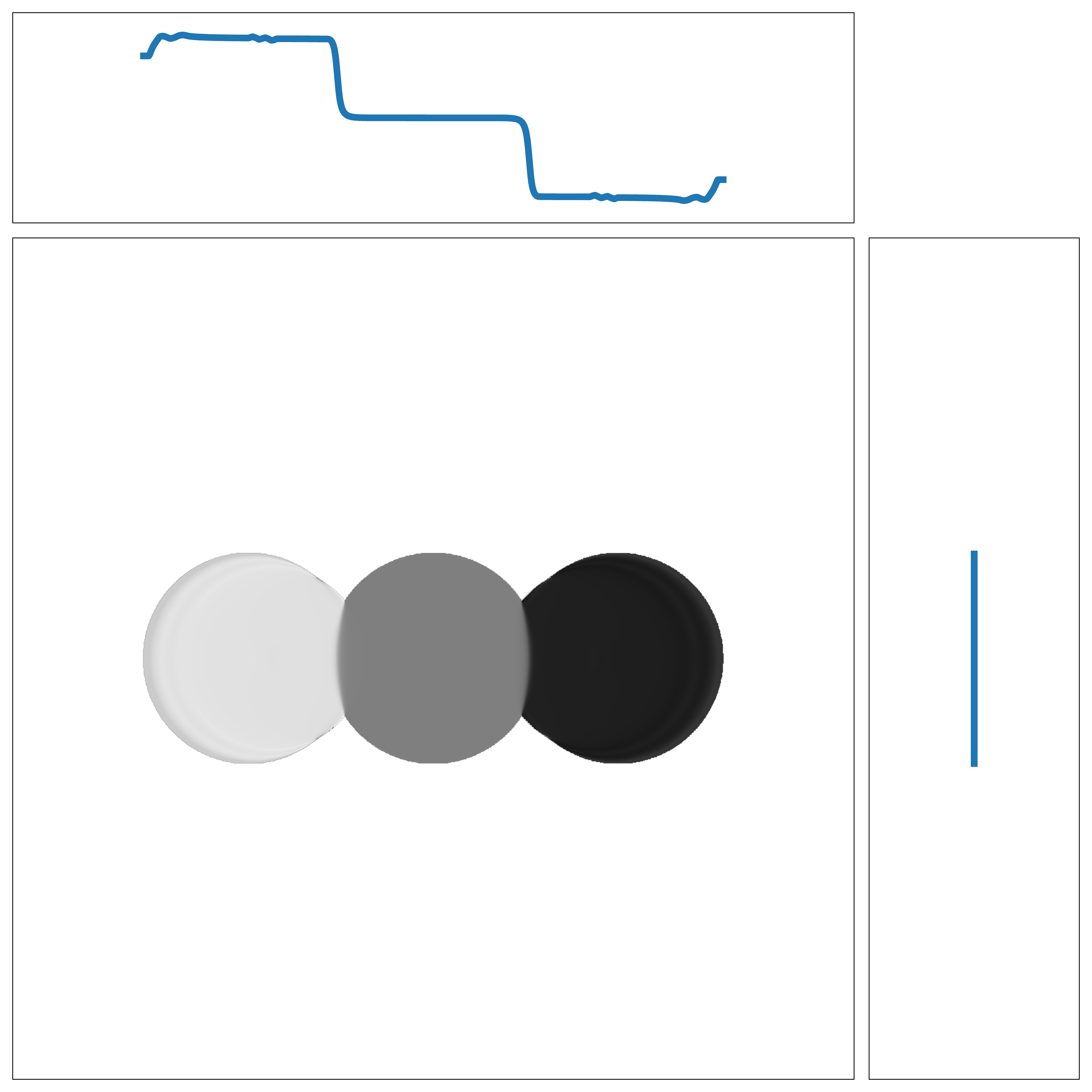} & Low power in high-freq nodes of MTF; PSF approximates incoherent overlapping of two single-aperture images   \\
\hline
\rule{0pt}{65pt} Transl. in x, 0.5 pix & \includegraphics[trim={3.7cm, 1.5cm, 3cm, 1.5cm}, clip=False, width=20mm, height=20mm]{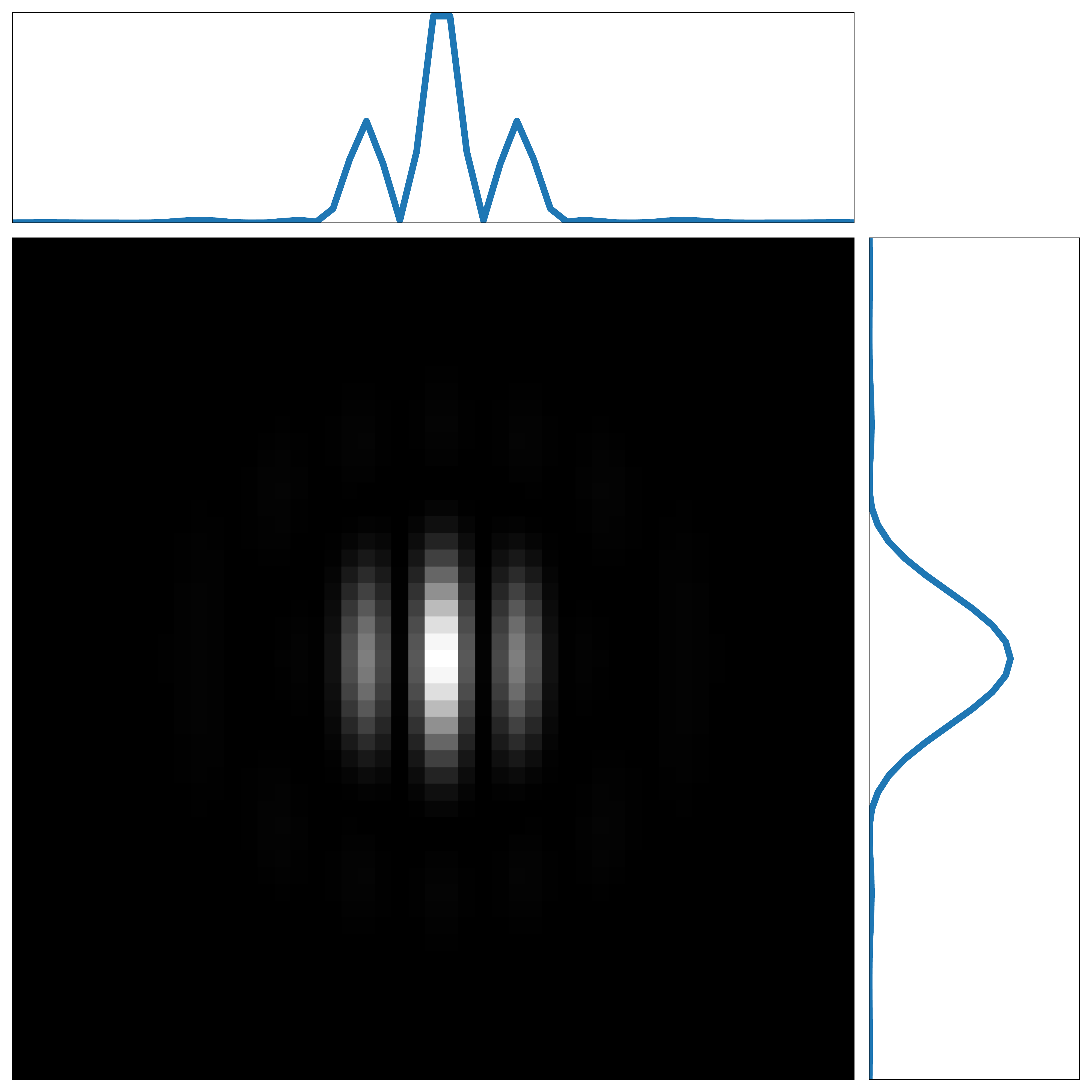} & \includegraphics[trim={3.7cm, 1.5cm, 3cm, 1.5cm}, clip=False, width=20mm, height=20mm]{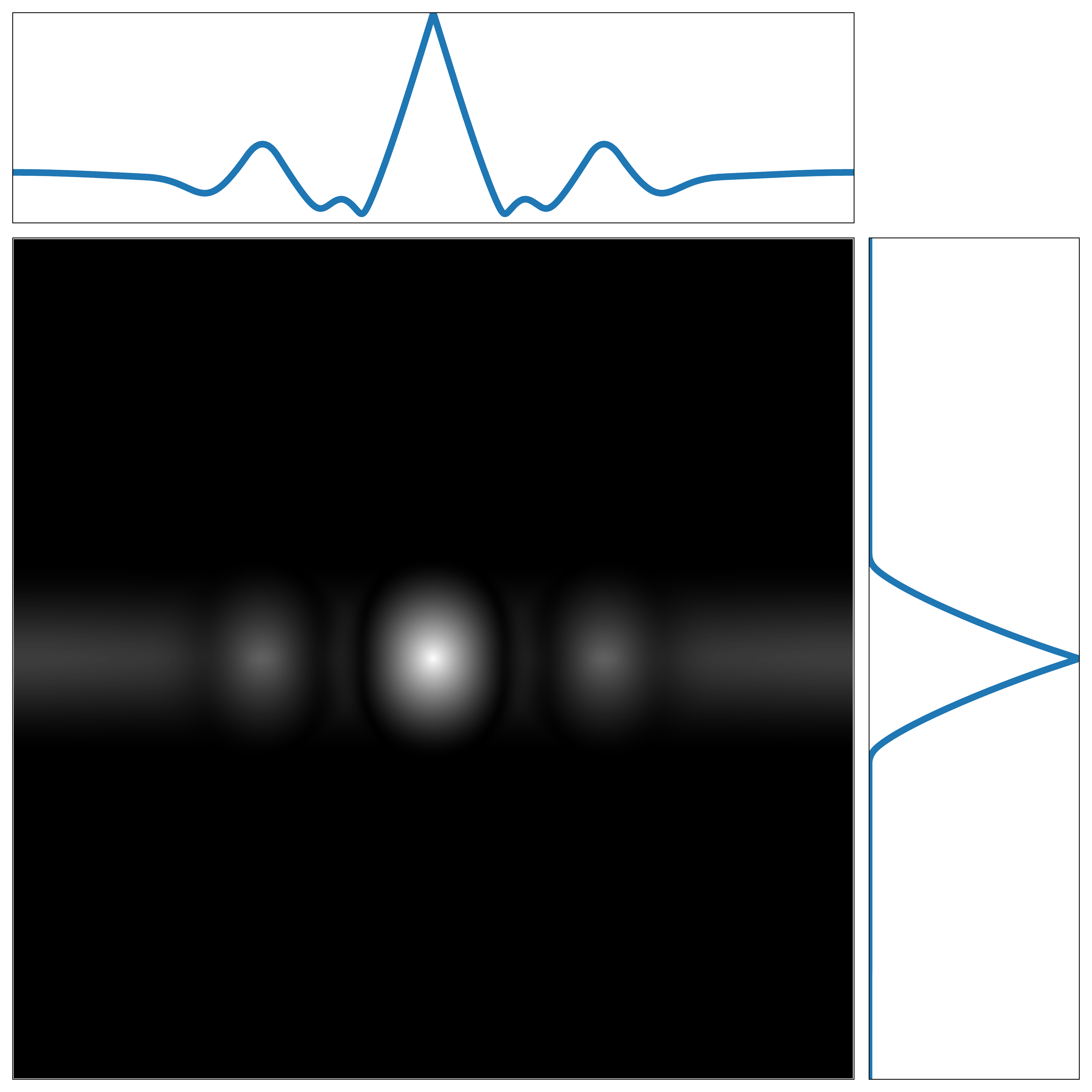} & \includegraphics[trim={3.7cm, 1.5cm, 3cm, 1.5cm}, clip=False, width=20mm, height=20mm]{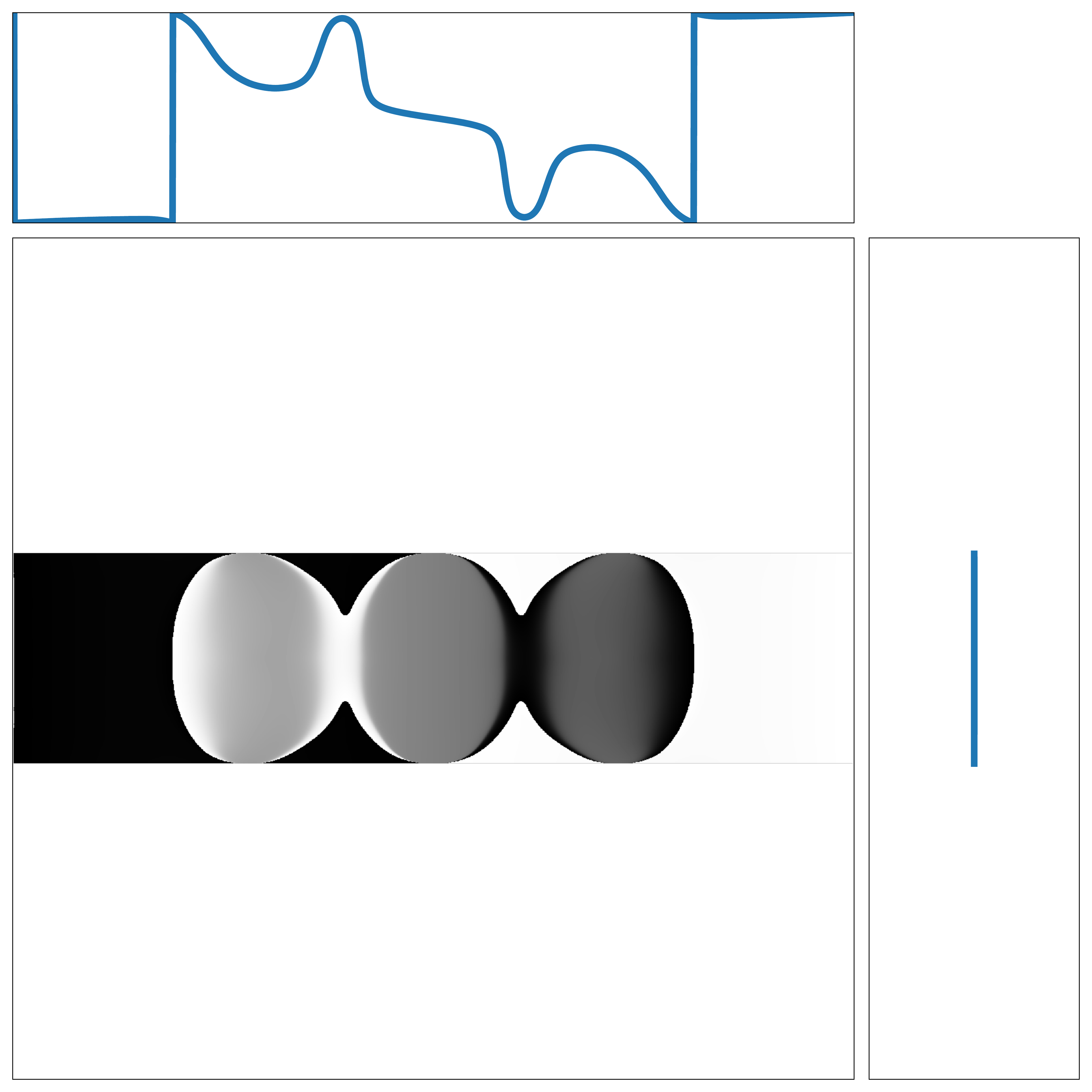} &  Compression of nodes in x, along with global phase gradient in x  \\
\hline
\rule{0pt}{65pt} Transl. in y, 2.0 pix  & \includegraphics[trim={3.7cm, 1.5cm, 3cm, 1.5cm}, clip=False, width=20mm, height=20mm]{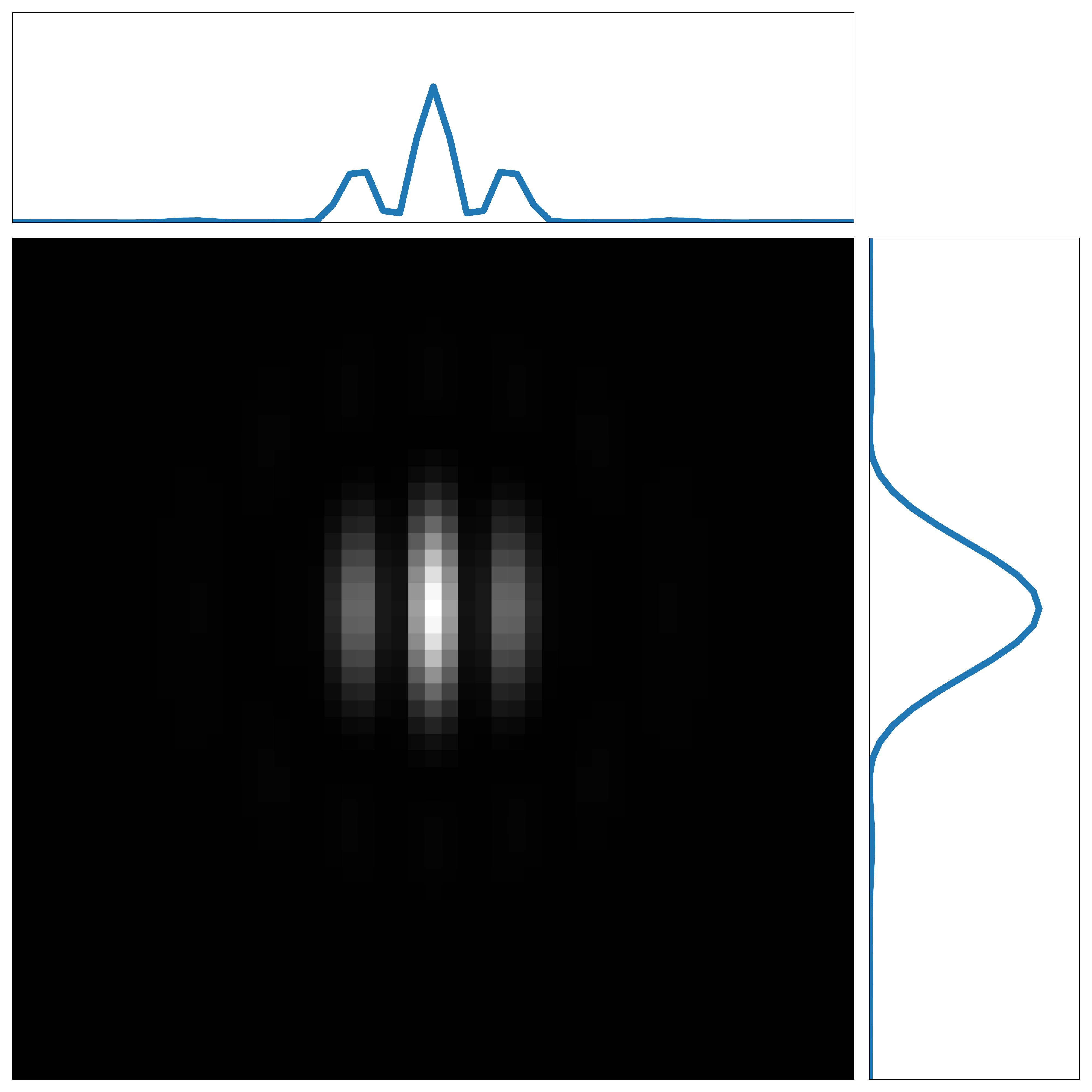} & \includegraphics[trim={3.7cm, 1.5cm, 3cm, 1.5cm}, clip=False, width=20mm, height=20mm]{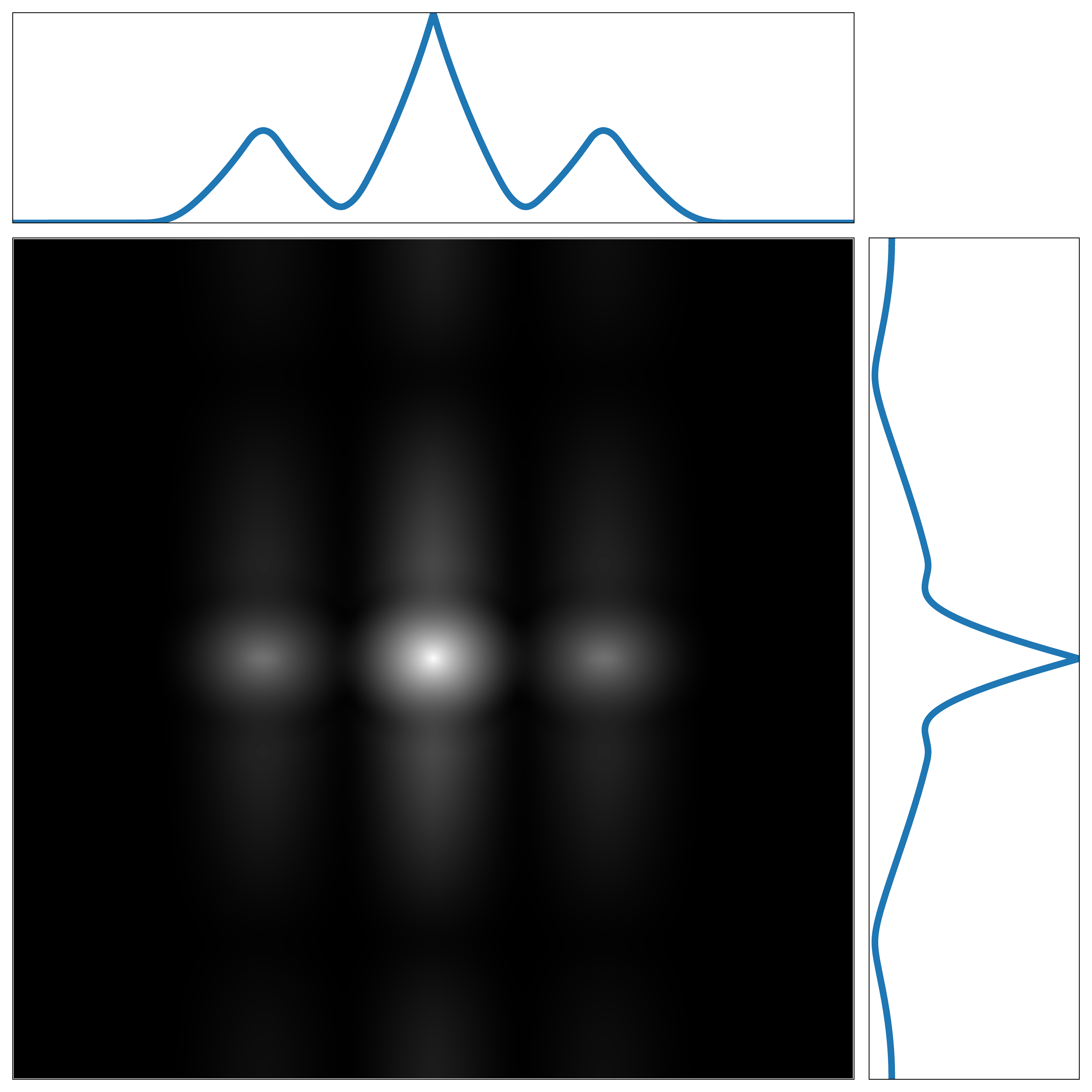} & \includegraphics[trim={3.7cm, 1.5cm, 3cm, 1.5cm}, clip=False, width=20mm, height=20mm]{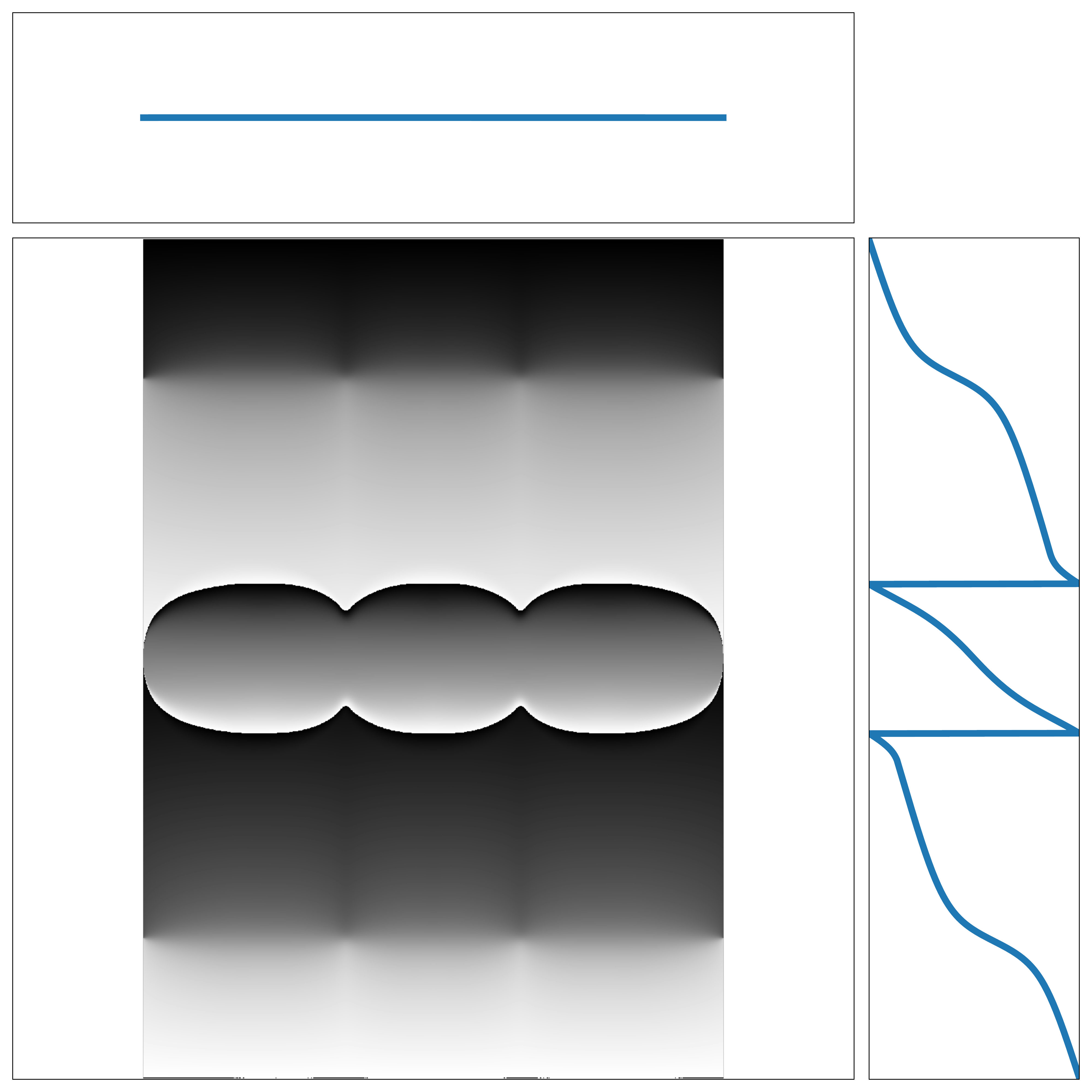} & Compression of nodes in y, along with global phase gradient in y  \\
\hline
\end{tabular}
\end{center}
\caption*{\small{PSF images are simulated with the LMIRcam sampling, i.e., 0.0107''/pixel. The blue lines in the 1D plots are horizontal and vertical cuts through the images. The 1D plots for the PSF images are normalized to the maximum of a `perfect' Fizeau PSF, whereas in the 1D plots for the MTFs, each plot is normalized with itself. Note the MTFs never go to zero between the nodes. The PTF 1D plot ranges are fixed to $\pm180^{\circ}$. Note some artifact aliasing appears in some images.}} \label{tab:sometab}
\noindent
\end{table}

\subsection{First on-sky phase-controlled Fizeau science observation}
\label{subsec:first_onsky}

\begin{figure}
\begin{center}
\includegraphics[height=8cm, trim={0.8cm 0 1.5cm 0}, clip=False]{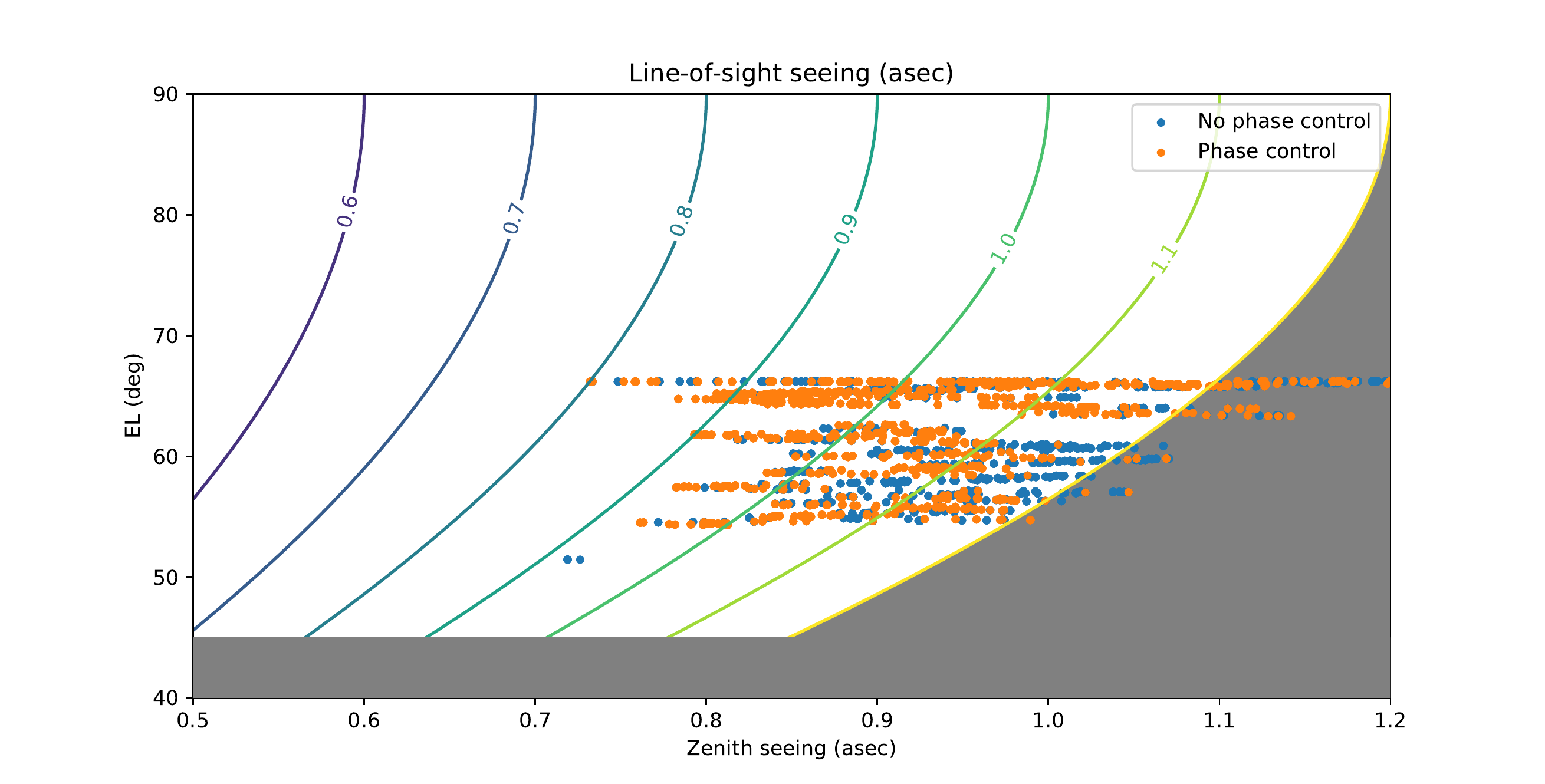}
\caption[]{Contoured region shows first-order limits on pointing in Fizeau mode, as imposed by line-of-sight seeing. Line-of-sight and zenith seeing are interconverted by a factor of $sin(EL)$. (Note that variations in EL tolerances and the validity of the $sin(EL)$ conversion can occur due to airflow and localized turbulence around the dome.) Favorable EL values of targets are within the contoured area. EL below 45 degrees is greyed out because the Phasecam loop becomes difficult to close below this threshold. The AO cannot perform at all below 30 degrees. The orange and blue points are from data during a Fizeau observation in May 2018 (see Sec. \ref{subsec:first_onsky}). ``Phase control'' is when Phasecam was actively controlling fringes. Note that some of the non-phase-controlled points were from during setup or alignment.}
\label{fig:good_el}
\end{center}
\end{figure}

In May 2018, Phasecam successfully phase-tracked a Fizeau observation of a bright star for up to a few minutes at a time, over a total range of two hours (Fig. \ref{fig:pcclosed_bool}). This was in seeing conditions that sometimes skirted the edge of the permissible range (Fig. \ref{fig:good_el}). Fig. \ref{fig:phys_psf} shows an example PSF from that observation, and Table \ref{table:fiz_open_closed} compares the PSF from that observation with and without Phasecam phase tracking over a period of a few seconds. The uncontrolled PSF exhibits OPD instability as well as phase smearing which has a detrimental effect on visibilities. The phase-controlled PSF is much more stable, though it occasionally suffers phase jumps equivalent to one wavelength in $K_{S}$-band. 

This observation was an important proof-of-principle that Fizeau with at least partial phase control can indeed be conducted for sustained periods with LBTI. In the process, the observing team gained a clearer picture of the technical exigencies of a more controlled Fizeau mode, including the updating of correction mirror setpoints, optimization of the Phasecam pupil illumination, counteraction of mirror hysteresis, and dealing with assorted complications such as the jumping away of the left-aperture component of the PSF in the focal plane when the Phasecam loop breaks. Reduction of the science dataset is ongoing.

\begin{figure}
	\begin{subfigure}{0.33\textwidth}
		\centering
		\includegraphics[height=5cm, trim={0.5cm 0 1.5cm 0}, clip=True]{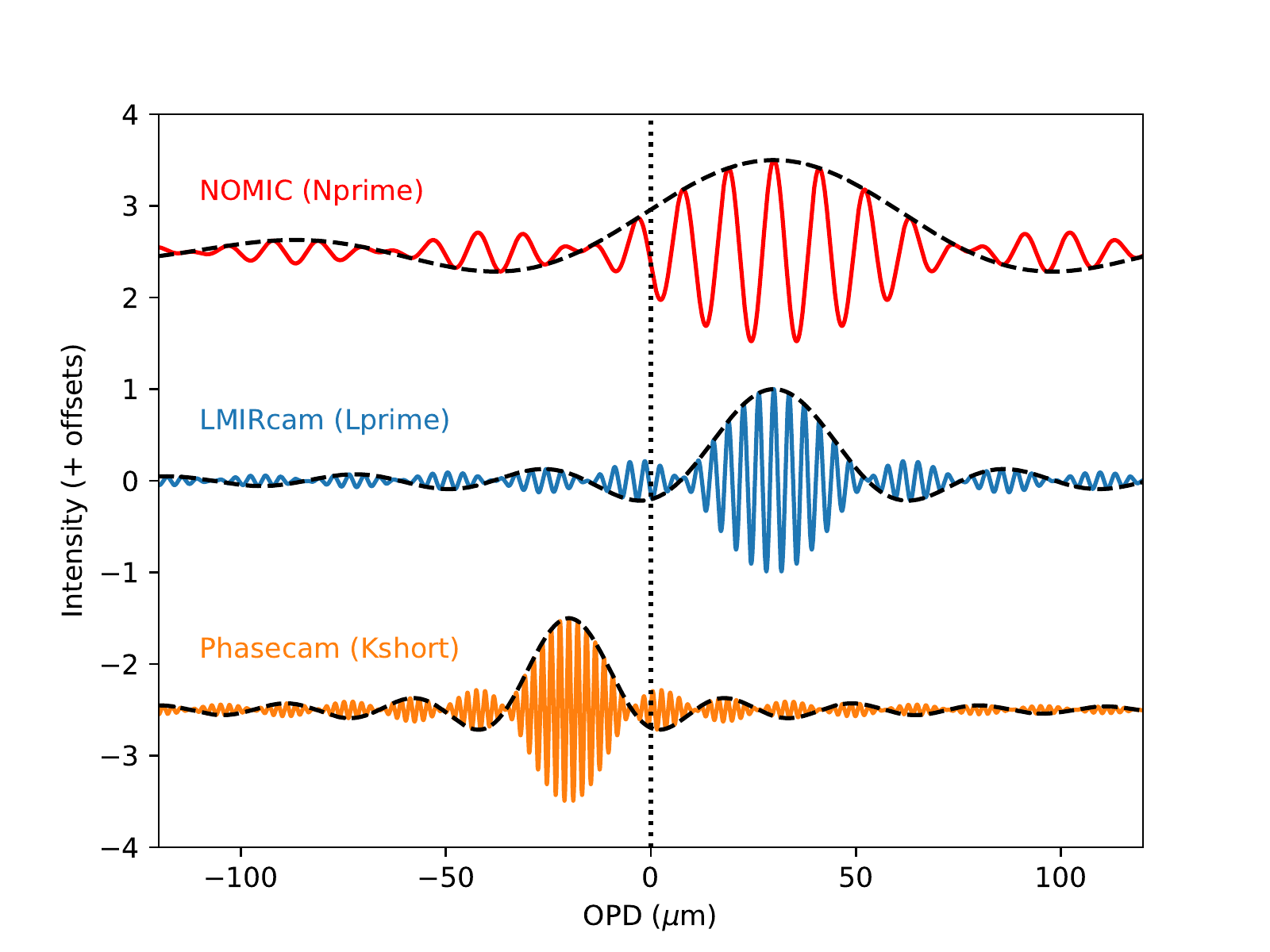}
		\caption{}
	\end{subfigure}\quad
	\begin{subfigure}{0.33\textwidth}
		\centering
		\includegraphics[height=5cm, trim={2.0cm 0 1.5cm 0}, clip=True]{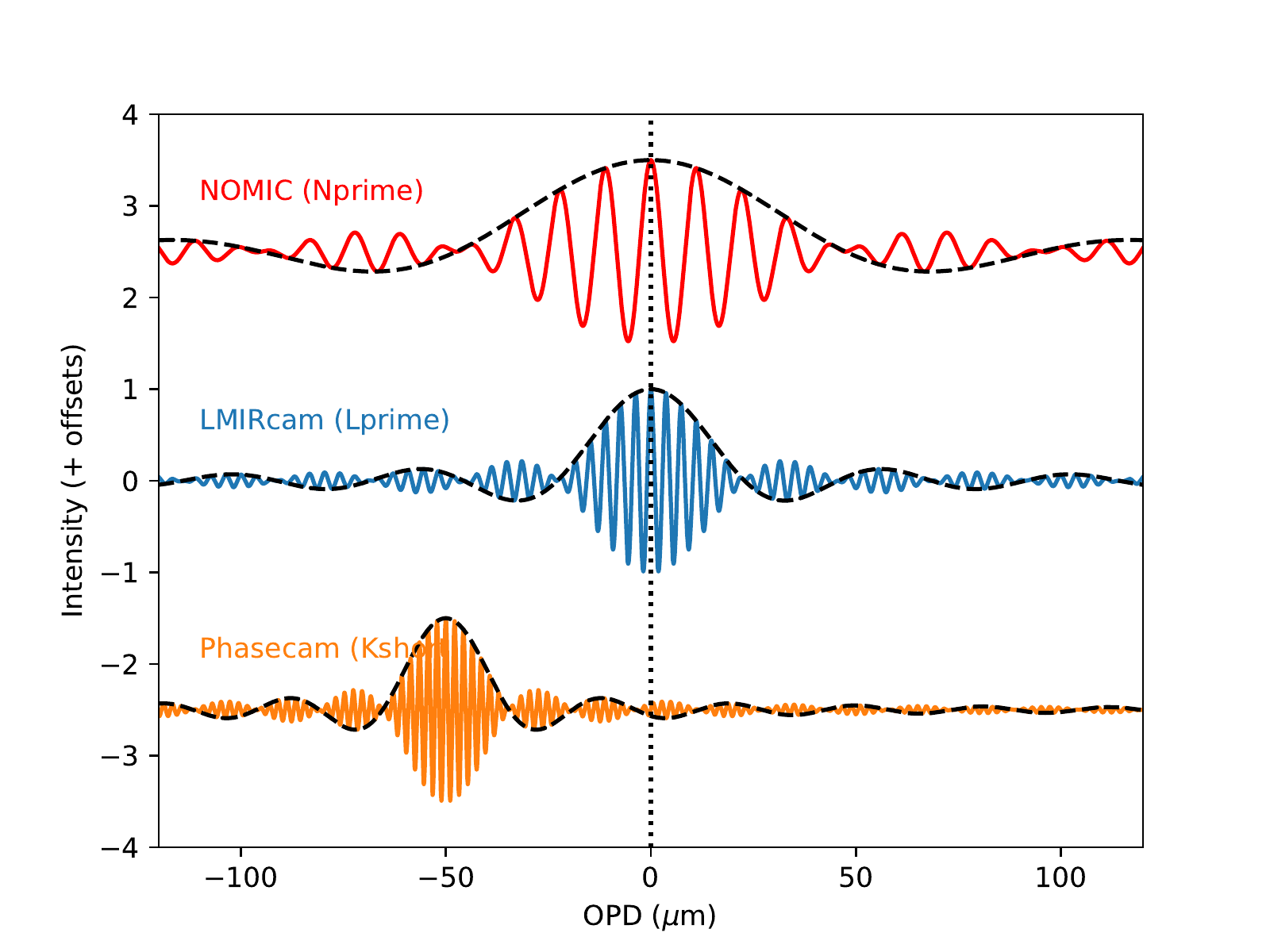}
		\caption{}
	\end{subfigure}\quad
	\hspace{-0.5cm}
	\begin{subfigure}{0.33\textwidth}
		\centering
		\includegraphics[height=5cm, trim={2.0cm 0 1.5cm 0}, clip=True]{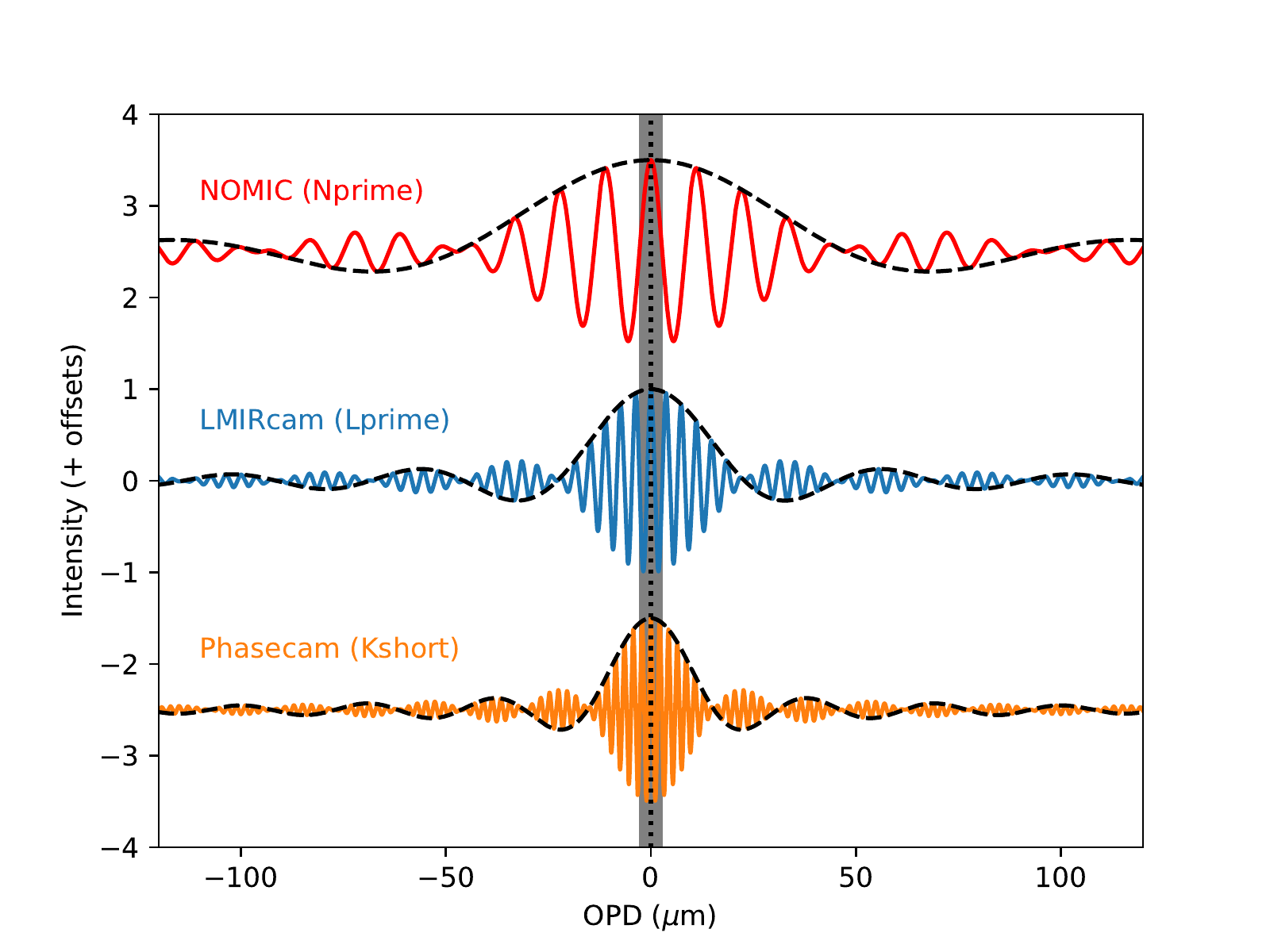}
		\caption{}
	\end{subfigure}\quad
\vspace{0.1cm}
\caption[]{A cartoon comparison of Fizeau-mode coherence envelopes on the science detectors and Phasecam at different stages of the OPD alignment. The modulation envelope due to the finite width of the filter bandpass is indicated with dashed lines. (Note that the NOMIC envelope here is only in NOMIC-Fizeau mode. In nulling mode, the Phasecam and NOMIC fringe envelopes share identical pathlengths by design.) After AO loop closure but before any mirror alignment. The coherence envelopes are at unknown OPD from the detector planes. b.) After inserting the grism, adjusting a pathlength corrector mirror, and acquiring and aligning the barber-pole fringes on either of the science detectors (see Fig. \ref{fig:barber}). This action also changes the OPD on Phasecam. c.) After adjusting the NIL beam combiner to center the coherence envelope on Phasecam. The NIL beam combiner is upstream of Phasecam and NOMIC-nulling, but not LMIRcam or NOMIC-Fizeau, so this will not change the OPD on the science detectors in Fizeau mode. The grey bar has width 5 $\mu$m to show the possible range of tolerance of the Phasecam loop before it opens.}
\label{fig:wave_packets}
\end{figure}

\begin{table}
\begin{center}
\caption{Current observing target requirements for performing controlled Fizeau} 
\label{table:observing_reqs}
\begin{tabular}{| l  | l | l |}
\hline
\textit{Parameter} & \textit{Requirement}  & \textit{Remarks} \\
\hline
DEC   & $\gtrsim -5^{\circ}$ & \begin{tabular}{@{}l@{}l@{}}Constrained by LBT latitude and \\ the need for $\leq$1.2'' seeing (see Fig. \ref{fig:good_el}) \end{tabular} \\
 \hline
Visibility & $\gtrsim$60-80\% in $K_{S}$-band & \begin{tabular}{@{}l@{}l@{}} Phasecam cannot lock onto extended\\sources.  \end{tabular} \\
 \hline
Observing wavelength & 
\begin{tabular}{@{}l@{}l@{}} $L$-, $M$-, or $N$-bands \\  (limited sensitivity in $K$-band) \end{tabular} & \begin{tabular}{@{}l@{}l@{}}Limited $K$-band is possible by reflecting\\some of this into Phasecam and some of\\it towards LMIRcam. \end{tabular} \\
 \hline
\begin{tabular}{@{}l@{}}$R$-band brightness \\ of AO guide star \end{tabular} & \begin{tabular}{@{}l@{}l@{}}$m_{R}\lesssim$9 mag (300 modes, 990 Hz)\\$9\lesssim m_{R}\lesssim 10.5$ (153 modes, 990 Hz)\\$10.5\lesssim m_{R}\lesssim 12.5$ (66 modes, 990 Hz)\\$12.5\lesssim m_{R}\lesssim 14.5$ (66 modes, 100s Hz)\\$14.5\lesssim m_{R}\lesssim 17.5$ (36 modes, $\approx$100 Hz)\end{tabular} & \begin{tabular}{@{}l@{}l@{}}For AO correction. Note AO guide stars\\have been acquired as far as $\sim$30''\\off-axis from the science target, but\\bright Fizeau targets will likely not\\require off-axis guide stars\end{tabular} \\
 \hline
$K$-band brightness & \begin{tabular}{@{}l@{}l@{}} $m_{K}\lesssim4.7$ for correction as slow\\as 520 Hz  \end{tabular}&     \begin{tabular}{@{}l@{}l@{}} For fringe tracking with Phasecam at \\standard detector binning. Slower\\cadences and adjustments in summer\\ 2018 may allow phase control of\\dimmer targets. (In principle it is\\possible to slow down the integration\\cadence until it dips below the\\atmospheric coherence timescale, at\\which point phase tracking becomes\\impossible. Fringe tracking can be\\neglected entirely at a steep cost in\\observing efficiency (see Sec. \ref{subsec:nascent}).) \end{tabular} \\
\hline
\end{tabular}
\noindent
\end{center}
\end{table}

\section{The future} 

During spring 2018 observations, we applied Phasecam phase control to the high-contrast Fizeau mode for the first time for any appreciable duration. This also gave the observing team experience for stabilizing the Fizeau mode in the future. Some of the current prerequisites for observing in Fizeau mode are summarized in Table \ref{table:observing_reqs}, and in the future these prerequisites may be loosened.

During the summer of 2018, we will work to construct a Fizeau calibration loop which uses science detector readouts to provide control that supplements the action of Phasecam. We have accordingly applied for time in fall 2018 to do on-sky testing. In the meantime, an upgrade of the left-side wavefront sensor was made in summer of 2018 as part of the SOUL project \cite{pinna2016soul}, after which the wavefront sensor is able to supply finer samplings, from 30$\times$30 across the pupil to a grid of 40$\times$40, and with a faster time response. Once both sides are upgraded, this will substantially improve the AO correction in mediocre weather conditions, meaning that there will be more opportunity for conducting imaging interferometry. We also hope to decrease the incidence of Phasecam's phase jumps by supplementing the $K_{S}$-band pupils with $H$-band. (See  \cite{maier2018phase} in these proceedings.) Reduction of already-captured Fizeau datasets will allow us to investigate achievable contrasts \cite{patru2017lbti1,patru2017lbti2} and make further improvements for making this observing mode routine.

\acknowledgments     
 
This material made use of CyVerse Atmosphere cloud computing, which is supported by the National Science Foundation under Grant Numbers \#DBI- 0735191 and \#DBI-1265383. URL: www.CyVerse.org.


\bibliography{report}   
\bibliographystyle{spiebib}   

\end{document}